\documentclass[a4paper,11pt]{article}

\usepackage[margin=1in]{geometry}
\usepackage{authblk}
\usepackage{subcaption}

\usepackage{amsmath}
\usepackage{amssymb}
\usepackage{amsthm}
\usepackage{graphicx}
\usepackage{epstopdf}
\usepackage{caption}
\usepackage[ruled,vlined]{algorithm2e}
\usepackage{bm} 
\usepackage{url} 

\title{Scale-Separated Dynamic Mode Decomposition and Ionospheric Forecasting}
\author[1,2,3]{Daniel J.  Alford-Lago\thanks{daniel.j.alford-lago.civ@us.navy.mil}}
\author[1]{Christopher W. Curtis}
\author[3]{Alexander T. Ihler}
\author[4]{Katherine A. Zawdie}
\affil[1]{Atmospheric Propagation Branch, Naval Information Warfare Center Pacific, San Diego, California, USA}
\affil[2]{Department of Mathematics and Statistics, San Diego State University, San Diego, California, USA}
\affil[3]{Department of Computer Science, UC Irvine, Irvine, California, USA}
\affil[4]{Space Science Division, Naval Research Laboratory, Washington, District of Columbia, USA}

\date{}
\begin{document}
\maketitle

\begin{abstract}
We present a method for forecasting the foF2 and hmF2 parameters using modal decompositions of ionospheric electron density profile time series. Our method is based on the Dynamic Mode Decomposition (DMD), which provides a means of determining spatiotemporal modes from measurements alone. DMD models are easily updated as new data is recorded and do not require any physics to inform the dynamics. However, in the case of ionospheric profiles, we find a wide range of oscillations, including some far above the diurnal frequency. Therefore, we propose nontrivial extensions to DMD using wavelet decompositions. We call this method the Scale-Separated Dynamic Mode Decomposition (SSDMD) since the wavelets isolate fluctuations at different time scales in the data into separated components. We show that this method provides a stable reconstruction of the peak plasma density and can be used to predict the state of foF2 and hmF2 at future time steps. We demonstrate the SSDMD method on data sets covering periods of high and low solar activity as well as low, mid, and high latitude locations.
\end{abstract}

\section{Introduction}\label{sec:intro}
The need for accurate modeling and forecasting of the prevailing space weather conditions continues to play a critical role in the development and operation of a variety of radio communications and radar applications. The Earth's ionosphere is of particular interest as it provides a medium for the propagation of radio waves far beyond the horizon \cite{ratcliffe_1959,  budden, davies_1990}. As a result, the ionosphere has been the subject of intense study for decades, and efforts to enhance our ability to model and predict the vertical plasma density profile continue to this day. Parameterizations of the height-dependent structure of the ionosphere include specifying the maximum plasma density value and the height at which it occurs. This peak in the plasma density profile is known as the F2-layer critical frequency, {\it foF2}, and is generally given in units of megahertz (MHz). The altitude at which the foF2 occurs is called {\it hmF2} and has units of kilometers (km). Together, these two parameters specify a crucial point in the local ionosphere that can have a considerable impact on radio propagation. Specifically, foF2 and hmF2 will affect the reflection height and thus ground distance that a radio wave at a given frequency will reach \cite{fagre_2019}. Therefore, misrepresenting the peak of the plasma density profile has immediate implications for military, commercial, and civilian applications. In general, there are two modeling approaches for ionospheric specification: physics-based and empirical.

In physics-based models, the equations of fluid mechanics and magnetohydrodynamics are solved. However, the ionosphere is driven by many exogenous systems, including solar and geomagnetic activity, tidal forcing from the lower troposphere \cite{liu_2016}, and thermospheric general circulation \cite{killeen_1987}. This means that while the physics are relatively well-understood, careful specification of these drivers is required in order to produce accurate simulations and forecasts. Additionally, even when physics-based models such as the thermosphere-ionosphere-mesosphere-electrodynamics general circulation model (TIME-GCM) \cite{dickinson_1981, roble_ridley, roble_1995} and SAMI3 \cite{huba_2000, huba_krall_2013} offer accurate modeling capability, they often underestimate the variance observed in the measurements of the ionospheric plasma density \cite{zawdie_2020}.   

On the other hand, empirical models, such as the International Reference Ionosphere (IRI), are generally less intensive to run but require large quantities of data from many different sources to account for the complex interactions between the various space weather systems. These sources include estimates from Mass Spectrometer Incoherent Scatter Radar (MSIS) to provide neutral composition derived from years of ground and space-based observations \cite{picone}, as well as vertical soundings for the bottomside, GPS-based observations of the total electron content (TEC), and {\it in situ} satellite measurements for the relevant ion species composition \cite{bilitza1}. Such an undertaking requires decades of dedicated service with international collaboration and has resulted in IRI becoming the official International Standardization Organization (ISO) standard for the ionosphere. Nevertheless, IRI provides only statistical estimates of the monthly average plasma density given several user-defined inputs such as solar activity via the monthly smoothed sunspot number and geomagnetic activity rather than simulating the dynamics.

More recently, determining reduced-order models (ROM) from data has been explored. In \cite{mehta_2018}, a quasi-physical dynamic ROM is obtained for the thermospheric mass density using the thermosphere-ionosphere-electrodynamics general circulation model (TIE-GCM) \cite{richmond_1992}, a precursor to TIME-GCM, as the source of observations. This ROM is based on a modal decomposition technique known as Dynamic Mode Decomposition (DMD) in which a set of spatiotemporal modes are determined via a linear best fit to data snapshots of a dynamical system \cite{schmid, mezic1, kutz}. DMD has also been shown to be especially useful in many physics and engineering contexts, such as in \cite{curtis} where it was used to help identify structure in weakly turbulent flows. Prior work on adapting DMD to data with dynamics at multiple scales can be found in \cite{dylewsky_2019, kutz4}, and building DMD models for nonlinear systems using deep learning in \cite{lago_2022}.

Our approach is motivated by the prevalence of vertical ionospheric sounder stations worldwide. These sounders generate data streams at regular cadences regarding the height-dependent profile of the ionospheric plasma density. However, plasma irregularities and traveling ionospheric disturbances manifest as fluctuations in the electron density profile (EDP) and occur over a range of time scales. Furthermore, the spatial frequencies of these irregularities are shown to range from the atmospheric scale height, where fluctuations are driven by gravity, down to the ion gyroradius, where fluctuations are driven by Earth's magnetic field \cite{booker_1978}.

We therefore see that modal analysis and dimensional reduction techniques, which facilitate the identification of simpler features within relatively complex data, would be of great utility in the study and use of ionospheric data. Likewise, measurement driven modeling techniques which bypass the intricate physics modeling that has been necessary to date to develop predictive capabilities would be especially desirable. To this end, we propose nontrivial extensions of DMD by way of wavelet decompositions that separate scales in a time series of EDPs. We call this method Scale-Separated DMD (SSDMD) and demonstrate its utility in obtaining a dynamic model of the local ionospheric peak density from a relatively short recording of data. 

SSDMD provides a novel approach to predicting the parameters foF2 and hmF2 that does not model their time evolution directly but instead uses the entire EDP time series to build a highly expressive model for the dynamics. Our key contribution is incorporating a wavelet decomposition and correlation analysis before applying DMD to the data. We find that critical couplings between scales that impact the stable evolution of DMD modes are preserved by grouping certain scales back together. These groupings are based on a one-step correlation that relates to how DMD is optimized. We find that the complete EDP forecasts from the method produce reasonable results in the F-region. However, the true utility of the method is the accuracy with which it predicts the foF2 and hmF2 parameters.

IRI was chosen for model comparison in this study because it is recognized as the official standard for the ionosphere by ISO, the International Union of Radio Science (URSI), the Committee on Space Research (COSPAR), and the European Cooperation for Space Standardization (ECCS) \cite{bilitza_2018}. While the number of ionospheric forecasting models seems to grow each year, we chose to use IRI as the gold standard because of its wide use in the community, \cite{bilitza1} having over 1,000 citations at the writing of this paper, and is accessible to the research community through simple programming APIs. While there are variants of IRI that employ more sophisticated techniques such as assimilation of real-time data \cite{galkin_2012}, these models are more complex and generally less accessible to the public. Moreover, the goal of this paper and the SSDMD model itself is not to outperform the most advanced, high-fidelity ionospheric models. Instead, we aim to provide a simple approach to forecasting key parameters using minimal amounts of data while providing reasonably accurate results that are on par with the most common and established methods.

This paper will provide the necessary background and algorithmic details to perform SSDMD on a time series of EDPs, and is organized as follows. In Section \ref{sec:dmd}, we present the DMD algorithm to compute spatial modes with time-evolving dynamics. Then, in Section \ref{sec:scalesep}, we demonstrate how we generate a scale-separated expansion of a signal using wavelet decompositions. Sections \ref{sec:corrgraphs} and \ref{sec:averages} then describe how we determine strong couplings across scales in the time series and average across them to produce an SSDMD model. Finally, Section \ref{sec:results} presents our results from this analysis on measured data from several Digisonde vertical sounders \cite{reinisch2011}.

\section{Method}\label{sec:method}
The SSDMD method presented here will generate a near-term, e.g., 48-hour, forecast of the local ionospheric conditions using a time series of EDPs from a vertical incidence sounder. In particular, we will use this model to generate a forecast of the peak plasma density, foF2, and height, hmF2. The method consists of four primary steps: 
\begin{enumerate}
    \item Use 1-dimensional wavelet decompositions at each fixed height in the data to separate fluctuations at different time scales and reconstruct the signal with each scale individually.
    \item Compute one-step correlations across the each scale reconstruction, determine which scales are strongly correlated, and add them together to form {\it connected components}.
    \item Average each connected component over 24-hour lags.
    \item Perform DMD on the averaged connected components to obtain a set of modes and eigenvalues for each.
\end{enumerate}
This algorithm will result in a separate DMD model for each of the averaged connected components. However, all these models will sum coherently to form a final reconstruction of the profile time series and predictions of its future state. From the forecasted profiles, we then compute the foF2 and hmF2 parameters. 

\begin{figure}[h!]
    \centering
    \includegraphics[width=0.95\textwidth]{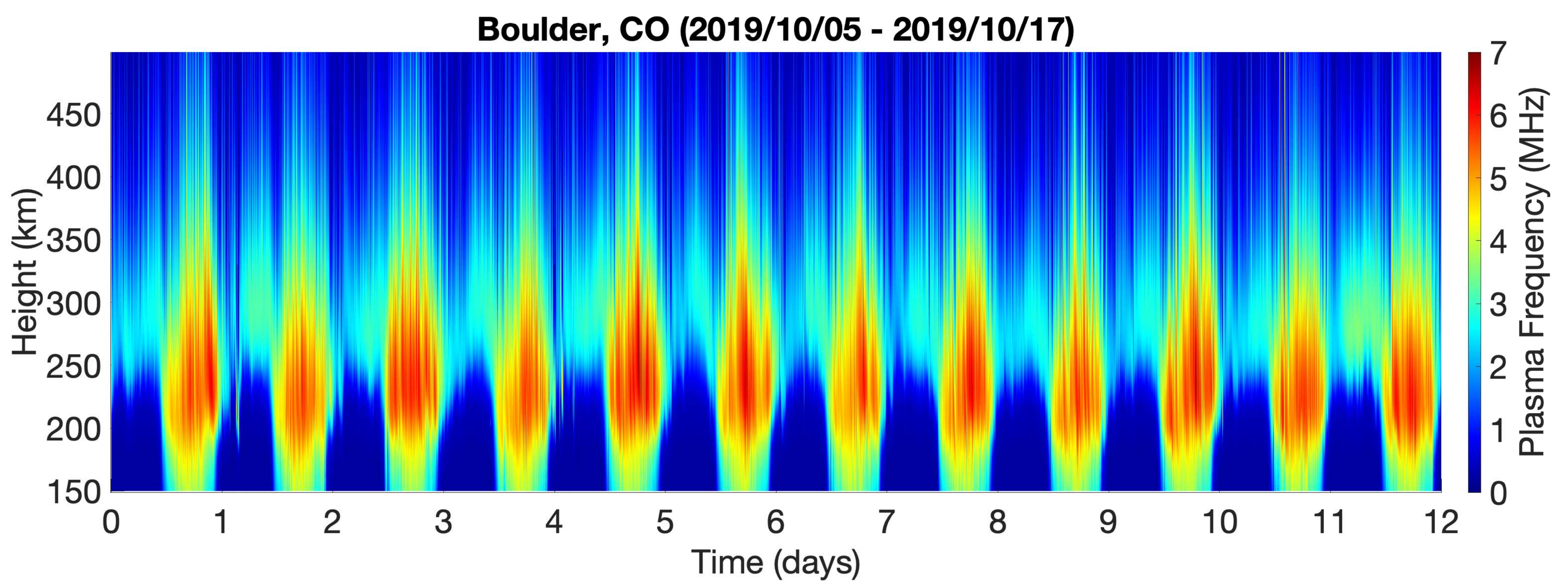}
    \caption{Dataset 1, a profilogram from the Digisonde Boulder, CO station covering the days of October 05, 2019 to October 17, 2019. Profiles were measured every 5 minutes.}
    \label{fig:datasets}
\end{figure}
The data used in this study are time series of ionospheric EDPs and their respective foF2 and hmF2 parameters gathered from two repositories, the Lowell GIRO Data Center digital ionogram database (Didbase) and the NOAA National Centers for Environmental Information (NCEI) Mirrion 2 data mirror. We will use a 12-day snippet, called Dataset 1, from a station in Boulder, Colorado, covering the dates 2019/October/05 to 2019/October/17 to illustrate each of the four steps of the SSDMD method above. This period of observation occurred near the last solar minimum yet still exhibits a wide spectrum of oscillations in the profile. 

Figure \ref{fig:datasets} shows Dataset 1 as a profilogram, which we have preprocessed by interpolating the raw sounder profiles to a regular 1km resolution height grid and then clipped below 150km. This is done because our model is intended to capture the dynamics of the F-layer parameters of the ionosphere. The following sections will now illustrate each step in SSDMD, starting with a description of the DMD method since it forms the basis of SSDMD.

\subsection{Dynamic Mode Decomposition}\label{sec:dmd}
DMD provides a method of finding a one-step, linear best-fit transformation from a time series of data that maps any observation in the series one time-step into the future. We start with a series of measurements of the system
\begin{equation}\label{eqn:ymatrix}
	{\bf Y} = \left\{ {\bf y}_{1} ~{\bf y}_{2} \cdots {\bf y}_{N_{\text{T}}}\right\},
\end{equation}
where ${\bf y}_{k} = {\bf y}(t_{k}) \in \mathbb{R}^{N_{S}}$ is a snapshot of the system at time $t_{k}$, thus ${\bf Y} \in \mathbb{R}^{N_{S}\times N_{T}}$. In the case of Dataset 1, each snapshot is a measurement of the vertical profile so each column in ${\bf Y}$ is an EDP. We assume a regular measurement cadence with $t_{k}=k \delta t$ for some time step $\delta t$, though in general this is not a requirement. From this, we create two new matrices
\begin{equation}\label{eqn:yplusminus}
	{\bf Y}_{-} = \left\{{\bf y}_{1} ~{\bf y}_{2} \cdots {\bf y}_{N_{T}-1} \right\} \quad \text{and} \quad {\bf Y}_{+} = \left\{{\bf y}_{2} ~{\bf y}_{3} \cdots {\bf y}_{N_{T}} \right\}
\end{equation}
and find a matrix ${\bf K}\in \mathbb{R}^{N_{S}\times N_{S}}$ such that
\begin{equation}\label{eqn:basicequation}
	{\bf K}{\bf Y}_{-} = {\bf Y}_{+}.
\end{equation}
This can be done simply via regression by solving the following optimization problem,
\begin{equation}\label{eqn:basicoptprob}
    {\bf K}_{o} = \underset{{\bf K}}{\mathrm{argmin}}\left|\left|{\bf Y}_{+} - {\bf K}{\bf Y}_{-}\right|\right|_{F}^{2}\,   = {\bf Y}_{+}{\bf Y}_{-}^{\dagger},
\end{equation}
where $\left|\left|\cdot\right|\right|_{F}$ denotes the Frobenius norm and ${\bf Y}^{\dagger}_{-}$ denotes the Moore-Penrose inverse of ${\bf Y}_{-}$. The DMD model is then given by the eigendecomposition of the matrix ${\bf K}_{o}$, however, solving \eqref{eqn:basicoptprob} directly can generate highly unstable results due to ill-conditioning in ${\bf Y}_{-}$. To address this, it is common in the DMD literature to use the singular-value decomposition (SVD) of ${\bf Y}_{-}$ and apply a threshold to keep only the most significant singular values. If the SVD of ${\bf Y}_{-}$ is 
\begin{equation}
	{\bf Y}_{-} = {\bf U}\mathbf{\Sigma}{\bf V}^{*},
\end{equation}
then introducing a threshold, $c_{\text{svd}}>0$, we truncate the columns of $\mathbf{U}$ and $\mathbf{V}$ corresponding to the singular values, $\Sigma_{jj}$, such that 
\begin{equation}\label{eqn:dmd_thresh}
	\text{log}_{10}\left(\frac{\Sigma_{jj}}{\Sigma_{11}} \right) > - c_{\text{svd}},
\end{equation}
where $\Sigma_{jj}$ are entries along the diagonal of $\mathbf{\Sigma}$ and are ordered such that 
\begin{equation}
	\Sigma_{11} \geq  \Sigma_{22} \geq \cdots \geq \Sigma_{N_{S}N_{S}}.
\end{equation}
We label the truncated versions of ${\bf U}$, $\mathbf{\Sigma}$, and ${\bf V}$ as $\tilde{{\bf U}}$, $\tilde{\mathbf{\Sigma}}$, and $\tilde{{\bf V}}$ respectively. A straightforward approximation of Equation \eqref{eqn:basicoptprob} can then be given by
\begin{equation}
	{\bf K}_{o} \approx {\bf Y}_{+}\tilde{{\bf V}}\tilde{\mathbf{\Sigma}}^{-1}\tilde{{\bf U}}^{*}.
\end{equation}
Note, ${\bf K}_{o}$ will be an $N_{S} \times N_{S}$ matrix, so when $N_{\text{S}}$ is very large it may be computationally expensive to compute the eigendecomposition; see \cite{tu1} for alternate formulations of DMD when this is the case. However, we found that the EDP data from a single sounding station is not high-dimensional enough to require these alternate forms. Instead, we simply compute the DMD modes and eigenvalues of ${\bf K}_{o}$ through the diagonalization
\begin{equation}
	\tilde{{\bf K}}_{o}={\bf W} \bm{\Lambda} {\bf W}^{-1}.
\end{equation}
where ${\bf W}$ is a matrix whose columns are eigenvectors, or DMD modes, and $\bm{\Lambda}$ is a diagonal matrix of DMD eigenvalues. For a given $\delta t$ representing the amount of time which has passed from observation ${\bf y}_{k}$ to ${\bf y}_{k+1}$, we construct a continuous-time model of the system,
\begin{equation}\label{eqn:dmdmodel}
	{\bf y}(t) \approx {\bf W} \bm{\Lambda}^{t / \delta t} {\bf W}^{\dagger} {\bf y}(0),
\end{equation}
where ${\bf y}(0)$ is some initial condition. Note that this decomposition provides a time stepping mechanism for reconstructing our time series that we may use for forecasting.

Comparisons of DMD to the well-established Empirical Orthogonal Function (EOF) analysis may be drawn. In practice, EOF models use Principal Component Analysis (PCA) to decompose the data into linear combinations of orthogonal functions. Fourier expansions of modulating coefficients for each component then provide variation over monthly and solar cycle scales; see \cite{liu_2008, zhang_2009, zhang_2013, mehta_2017, li_2021} for in-depth description of EOF analysis for space weather. This has the advantage of including proxies for external drivers such as the F10.7-cm solar flux in the forecast. Nevertheless, such indices are not readily available on the time scales that we are able to measure ionospheric profiles and provide little additional input for a 24- to 48-hour forecast.  Furthermore, EOF models are restricted to an orthogonal basis of functions for the dynamics due to the use of PCA. The DMD modes have no such restriction since they are derived from the eigendecomposition of the ${\bf K}_o$ matrix. Another major difference between our method and conventional EOF models is we separate the various time scales in the data prior to fitting the DMD modes and eigenvalues.

Thus, beyond just producing a modal decomposition from data, the DMD method gives a time-evolving model for said data through the spectra of the ${\bf K}$ matrix. Further connections between DMD and dynamical systems analysis can be established through its relationship with the Koopman operator \cite{koopman}; see Appendix \ref{sec:koopman}. While a generally successful approach, this straightforward implementation struggles with multiscale data or any data that has both very small and very large gradients from snapshot to snapshot due to the one-step regression in Equation \ref{eqn:basicoptprob}. This motivates the use of some form of temporal scale separation.

\subsection{Scale Separation of EDP Time Series}\label{sec:scalesep}
\begin{figure}[h!]
    \centering
    \includegraphics[width=0.95\textwidth]{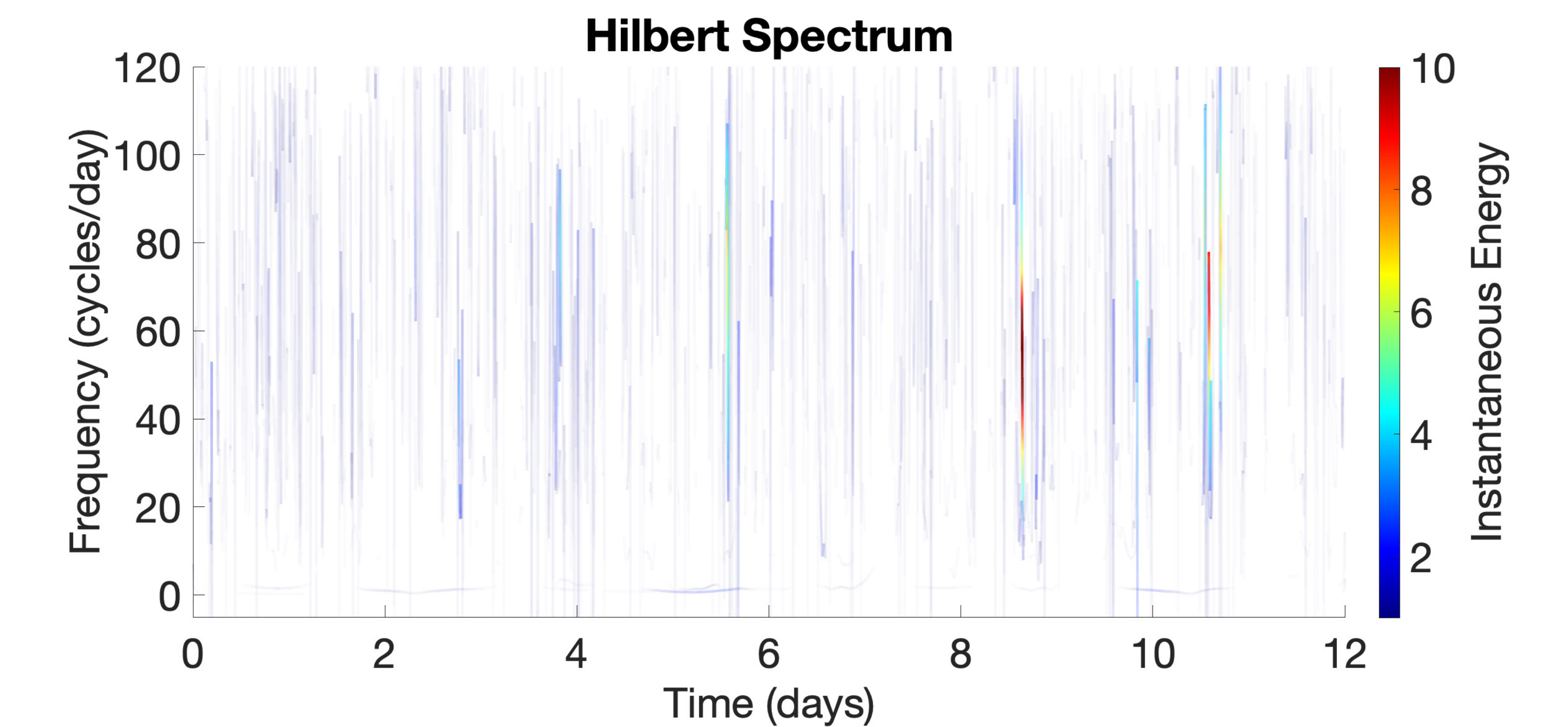}
    \caption{The affiliated Hilbert spectrum for a slice through Dataset 1 at a height of $400$km. The Hilbert spectrum plot reveals the instantaneous energy in the data as a function of time and frequency. The stable diurnal oscillation can be see near 1 cycle/day, while various time localized, spurious oscillations occur throughout at frequencies that are an order of magnitude higher.}
    \label{fig:hilbert}
\end{figure}

The primary contribution of this paper is to provide a method of adapting the DMD algorithm to work on data with fluctuations at multiple scales, as is the case when modeling EDP measurements. The need to account for these oscillations is motivated by the Hilbert spectrum of a slice through Dataset 1 at a vertical height of $400$km. At this altitude, we see there is a significant degree of instantaneous energy at frequencies much higher than diurnal variation (1 cycle/day); see Figure \ref{fig:hilbert}. These relatively high-frequency, transient events complicate direct applications of DMD, but do not necessarily represent noise that should be filtered out.

We therefore use a multiresolution analysis by way of 1-dimensional wavelet decompositions to facilitate DMD; see \cite{mallat, mallat_1989} for in-depth theory and applications of wavelet decompositions. For a given time series ${\bf y}(t) \in \mathbb{R}^{N_{s}}$ representing vector observations of EDPs, we decompose each height in the time series into $N_{lvl}$ levels, such that 
\begin{equation}\label{eqn:wavedec1}
	{\bf y}(t) \approx \sum_{j=1}^{N_{lvl}+1}{\bf d}_{j}(t),
\end{equation}
where ${\bf d}_{j}(t) \in \mathbb{R}^{N_{s}}$, such that
\begin{equation}\label{eqn:wavedec2}
	{\bf d}_{j}(t) = \sum_{n=-M_{f}}^{M_{f}} {\bf d}_{j,n}  \psi_{j,n}(t), ~ 1\leq j \leq N_{lvl},
\end{equation}
and
\begin{equation}\label{eqn:wavedec3}
	{\bf d}_{N_{lvl}+1}(t) = \sum_{n=-M_{f}}^{M_{f}}{\bf d}_{N_{lvl}+1,n} \phi_{N_{lvl},n}(t),
\end{equation}
where $\psi(t)$ and $\phi(t)$ are the wavelet and scaling functions of the decomposition, respectively,
\begin{align}\label{eqn:wavedec4}
	\psi_{j,n}(t) &= 2^{-j/2}\tau_{n}\psi\left(2^{-j}t \right), \nonumber\\
    \phi_{N_{lvl},n}(t) &= 2^{-N_{lvl}/2}\tau_{n}\phi\left(2^{-N_{lvl}}t \right).
\end{align}
The vectors ${\bf d}_{j,n}$, $1\leq j \leq N_{lvl}$, denote the {\it detail coefficients} at the $j^{th}$ scale while ${\bf d}_{N_{lvl}+1,n}$ denotes the {\it approximation coefficients} at the terminal scale. 

With the wavelet decompositions performed independently at each height in the profile, the vector quantities ${\bf d}_{j}(t)$ represent only parts of the signal at the $j^{th}$ scale at time $t$. Given our discrete time series from Equation \eqref{eqn:ymatrix}, these vector quantities form the columns of a new set of data matrices,
\begin{equation}
	{\bf Y}_{j} = \left\{{\bf d}_{j,1}~{\bf d}_{j,2}~\cdots~{\bf d}_{j, N_{T}} \right\},
\end{equation}
which are reconstructions of the original data at each scale and sum coherently, so that ${\bf Y} = \sum_{j=1}^{N_{lvl}+1} {\bf Y}_{j}$.

\begin{figure}[h!]
    \centering
    \includegraphics[width=1\textwidth]{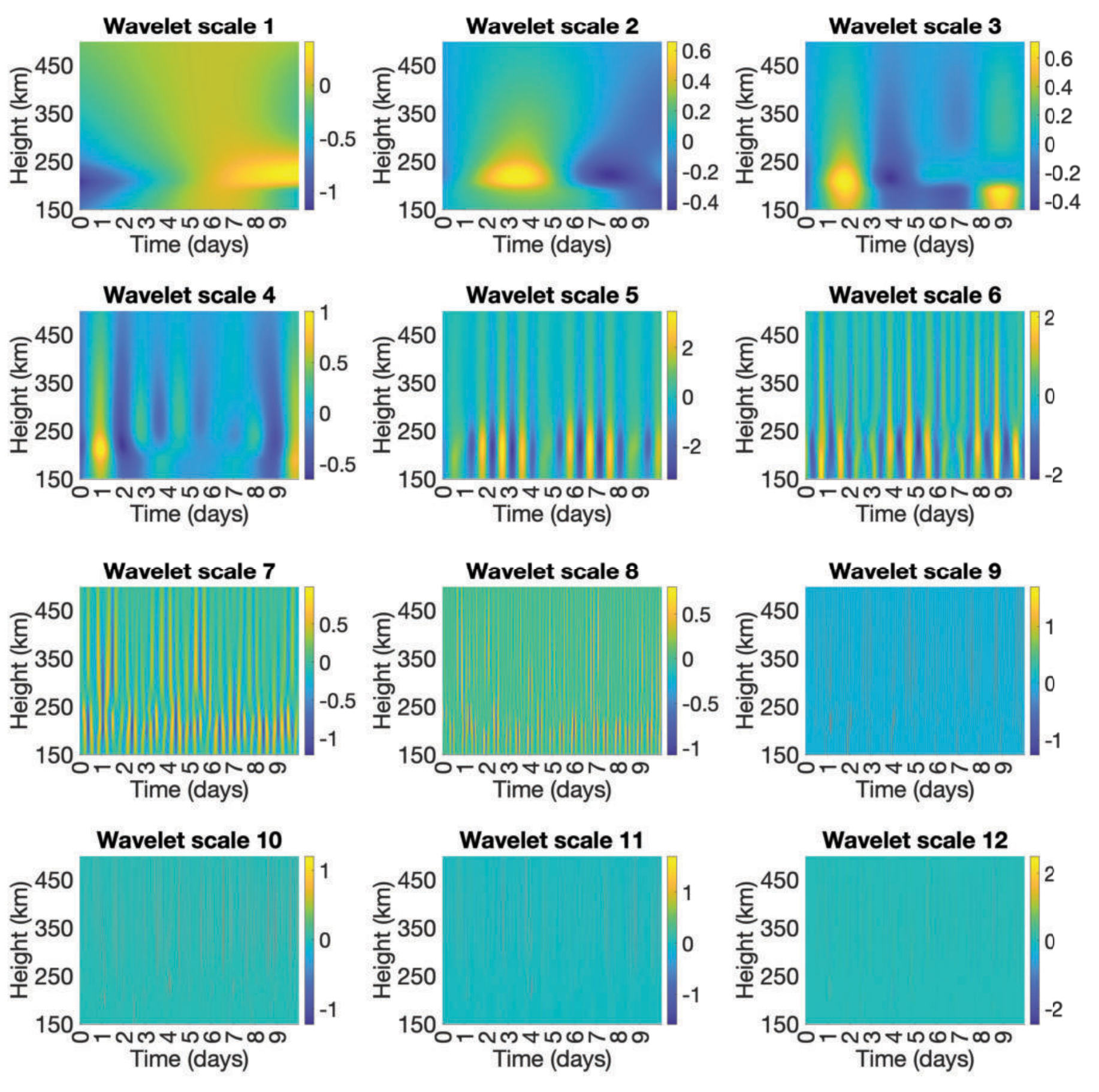}
    \caption{Dataset 1 decomposed into 12 scales. Each panel is a reconstruction of the full EDP time series using only the $j^{th}$ scale coefficients from the wavelet decomposition at each height, ${\bf Y}_j$. The color axis represents plasma frequency in MHz.}
    \label{fig:wave_scale_recons_2019}
\end{figure}
In Figure \ref{fig:wave_scale_recons_2019}, we have Dataset 1 expanded into 12 scale reconstructions. These scales further illustrate the multiscale nature of high-resolution EDP measurements, with fluctuations on the order of 1-2MHz in magnitude observed up to the fastest scales. These sub-diurnal oscillations can appear as broad-spectrum noise in the raw profilogram and can make modal decompositions like DMD quite challenging. Note that the diurnal oscillation itself does not appear until the $5^{th}$ or $6^{th}$ scale in Figure \ref{fig:wave_scale_recons_2019}, and several longer-period trends are observed before the terminal scale. In the following section we will see how these oscillations can be highly correlated in terms of an optimal DMD one-step fit. Fourth-order Coiflets were used for the discrete wavelet transforms. The wavelet type is a model hyperparameter and may vary for different data sets. However, we found that this choice worked well for all test cases in this study.

\subsection{Computing Correlations Across Scales}\label{sec:corrgraphs}
Applying DMD to each scale separately does not produce optimal results and can even produce DMD modes that are unstable and decay to zero or grow to infinity almost immediately. Instead, we found correlations across each of the scales can indicate strong dynamical couplings between them, and preserving these has a pronounced impact on the fidelity and stability of the DMD modes. Identifying the strength of these couplings required developing a measure of correlation that takes into account the role that the matrix ${\bf K}_{o}$ plays in advancing the data forward in time. To this end, we defined the following correlation matrix ${\bf C}$ whose entries are given by
\begin{equation}\label{eqn:corrcoeff}
	C_{jl}  = \left[ \overline{\tilde{{\bf Y}}_{j,+} \odot \tilde{{\bf Y}}_{l,-}} + \overline{\tilde{{\bf Y}}_{j,-} \odot \tilde{{\bf Y}}_{l,+}}\right],
\end{equation}
with, $j,l \in {1,\dots, N_{lvl}}$, and 
\begin{equation}\label{eqn:varscale}
	\tilde{{\bf Y}}_{j} = \frac{{\bf Y}_{j} - \overline{{\bf Y}}_{j}}{\left|\left|{\bf Y}_{j} - \overline{{\bf Y}}_{j} \right|\right|_{2,t}}.
\end{equation}
The $\overline{\cdot}$ and $[\cdot]$ denote taking the mean in the time and space dimensions of the time series, respectively, $\left|\left|\cdot\right|\right|_{2,t}$ is an $L_{2}$-norm over time, and $\odot$ is the Hadamard product between two matrices. Finally, the $+$ and $-$ subscripts indicate shifting the time series forward or backward one time step as in Equation \ref{eqn:yplusminus}.

Because ${\bf K}_o$ is optimized to advance any profile in the data one time step into the future, this correlation coefficient provides a quantitative means for comparing the time series across different timescales in the context of fitting optimal DMD modes. Then, by setting a threshold value, $c_{\text{corr}}$, we generate an adjacency matrix $\mathbf{A}$ with entries 
\begin{equation}\label{eqn:corr_thresh}
A_{jl} = \left\{
\begin{array}{rl}
1, & |C_{jl}| \geq c_{\text{corr}} \\ 
0, & |C_{jl}| < c_{\text{corr}}
\end{array}
\right.
\end{equation}
The matrix $\mathbf{C}$ is symmetric, and so ${\bf A}$ is as well. Note, in practice these correlations will typically be larger for the longer time scales since we are looking at one-step correlations, with higher frequency oscillations becoming increasingly less correlated. The matrix ${\bf A}$ generates a graph $G$ that indicates which of the ${\bf Y}_{j}$ scale reconstructions should be grouped back together to preserve their dynamic coupling.

\begin{figure}[ht]
    \centering
    \includegraphics[width=1\textwidth]{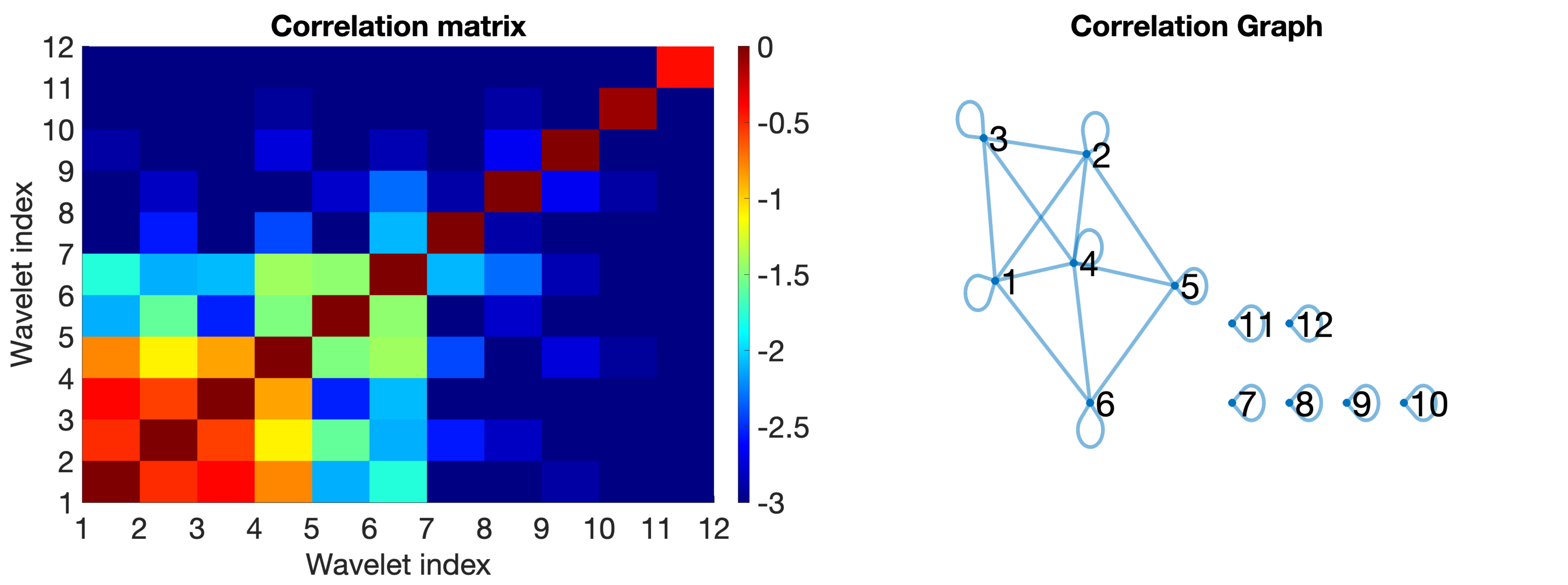}
    \caption{The correlation coefficient matrix ${\bf C}$ (left) and the corresponding graph $G$ (right) indicating which scales are highly coupled. The correlation threshold $c_{corr}=-1.95$ was used for Dataset 1.}
    \label{fig:correls}
\end{figure}

Thus, for a given choice of threshold $c_{\text{corr}}$, we will have $N_{C}\leq N_{lvl}+1$ connected components within $G$. We then form $N_{C}$ new time series by summing only the ${\bf Y}_{j}$ which belong to the same connected component,
\begin{equation}
	{\bf Y}_{n}^{\text{C}} = \sum_{j \in G_{n}} {\bf Y}_{j},
\end{equation}
where $j \in G_{n}$ denotes the scales that are in the $n^{th}$ connected component in $G$, and ${\bf Y}^{\text{C}}_{n}$ is the time series for the $n^{th}$ connected component. Figure \ref{fig:correls} shows the matrix ${\bf C}$ and the graph $G$ for Dataset 1. Note that the first group consists of the bulk of the large scale features in the time series while the higher frequency scales remain on their own. However, this may not always be the case, and subgroups within the high frequency components could arise depending on the data observed.
\begin{figure}[h!]
    \centering
    \includegraphics[width=1\textwidth]{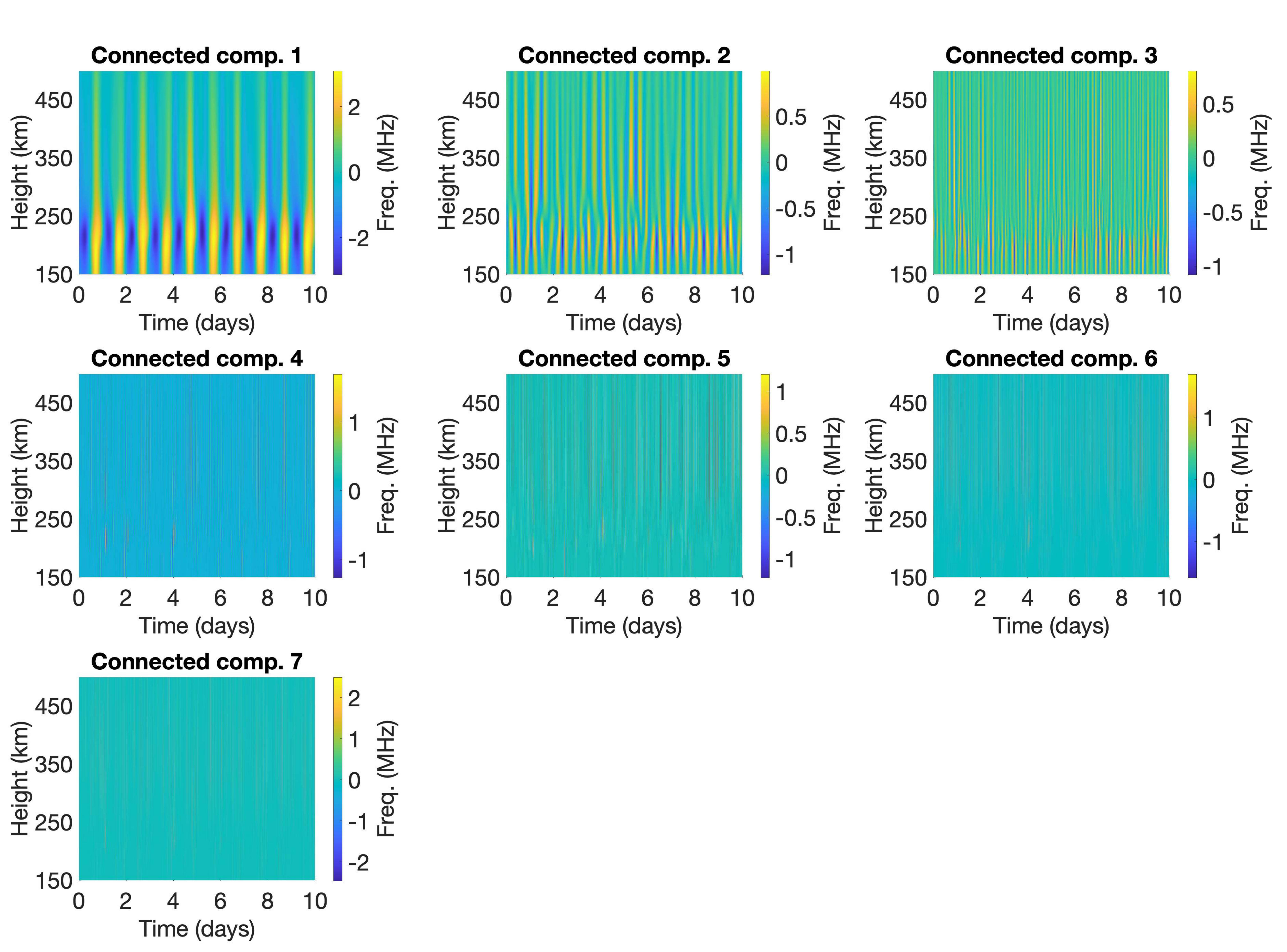}
    \caption{Dataset 1 decomposed into 7 connected components. Each component captures features of the data with strong correlations according to the one-step spatiotemporal coefficients.}
    \label{fig:wave_group_decomp_2013}
\end{figure}

At this point, one could find a corresponding $\tilde{\mathbf{K}}_{o,n}$ via DMD and generate an affiliated expansion for each connected component so that the total time series can be approximated by
\begin{equation}
	{\bf y}(t) \approx \sum_{n=1}^{N_{C}} {\bf W}_{n} \bm{\Lambda}^{t / \Delta t}_{n} {\bf W}^{\dagger}_{n} {\bf y}_{n,0}.
\end{equation}
However, we note that, using observations that span only several days in time, the EDP at a single sounding station is essentially memoryless after twenty-four hours have passed \cite{araujo_2005}. This strongly suggests that before naively applying the DMD method to time series of arbitrary length, instead, we should first average the data across 24-hour cycles for the duration of our measurement period. 

\subsection{Averaging for DMD}\label{sec:averages}
Having decomposed the EDP time series into correlated time scales, we now have a collection of time series, 
\begin{equation}
	{\bf Y}_{1}^{\text{C}}, ~ {\bf Y}_{2}^{\text{C}}, ~ \cdots, ~ {\bf Y}_{N_{C}}^{\text{C}},
\end{equation}
that represent scales within the data set whose one-step correlations are relatively weak. We treat these as being essentially independent with respect to our DMD approximation.

Denoting the number of time steps in a full day as $T_{D}$ and assuming that $N_{T}+1$ is divisible by $T_{D}$, so that the data set represents the number of days $N_{D}$ where
\begin{equation}
	N_{D} = \frac{N_{T}+1}{T_{D}},
\end{equation}
we isolate the mean signal over 24-hour cycles from the fluctuations about the mean for each ${\bf Y}_{n}^{\text{C}}$. This creates two new affiliated time series for each connected component that have the properties,
\begin{equation}
	\bar{{\bf y}}_{n}^{\text{C}}(t_{k} + T_{D}) = \bar{{\bf y}}_{n}^{\text{C}}(t_{k}),
\end{equation}
and 
\begin{equation}\label{eqn:meancycles}
	\sum_{k=1}^{T_{D}} \hat{{\bf y}}_{n}^{\text{C}}(t_{k} + mT_{D}) = 0, \quad m=0,\dots,N_{D}-1,
\end{equation}
where $\bar{\cdot}$ and $\hat{\cdot}$ denote the 24-hour mean signal and fluctuations about the 24-hour mean, respectively. The fluctuations in Equation \ref{eqn:meancycles} effectively represent the noise signal for each component. These may prove useful in future experiments to generate nonparametric error estimates, however, in this paper they are not used further since our goal is to forecast parameters derived from the profile. Taking the vector quantities, $\bar{{\bf y}}_{n}^{\text{C}}$ to be columns of new mean-signal matrices we have
\begin{equation}
	\bar{{\bf Y}}_{n}^{\text{C}} = \left\{\bar{{\bf y}}_{n,1}^{\text{C}},~\bar{{\bf y}}_{n,2}^{\text{C}},~\dots,~\bar{{\bf y}}_{n,T_{D}}^{\text{C}}\right\}.
\end{equation}

\begin{figure}[h!]
    \centering
    \includegraphics[width=1\textwidth]{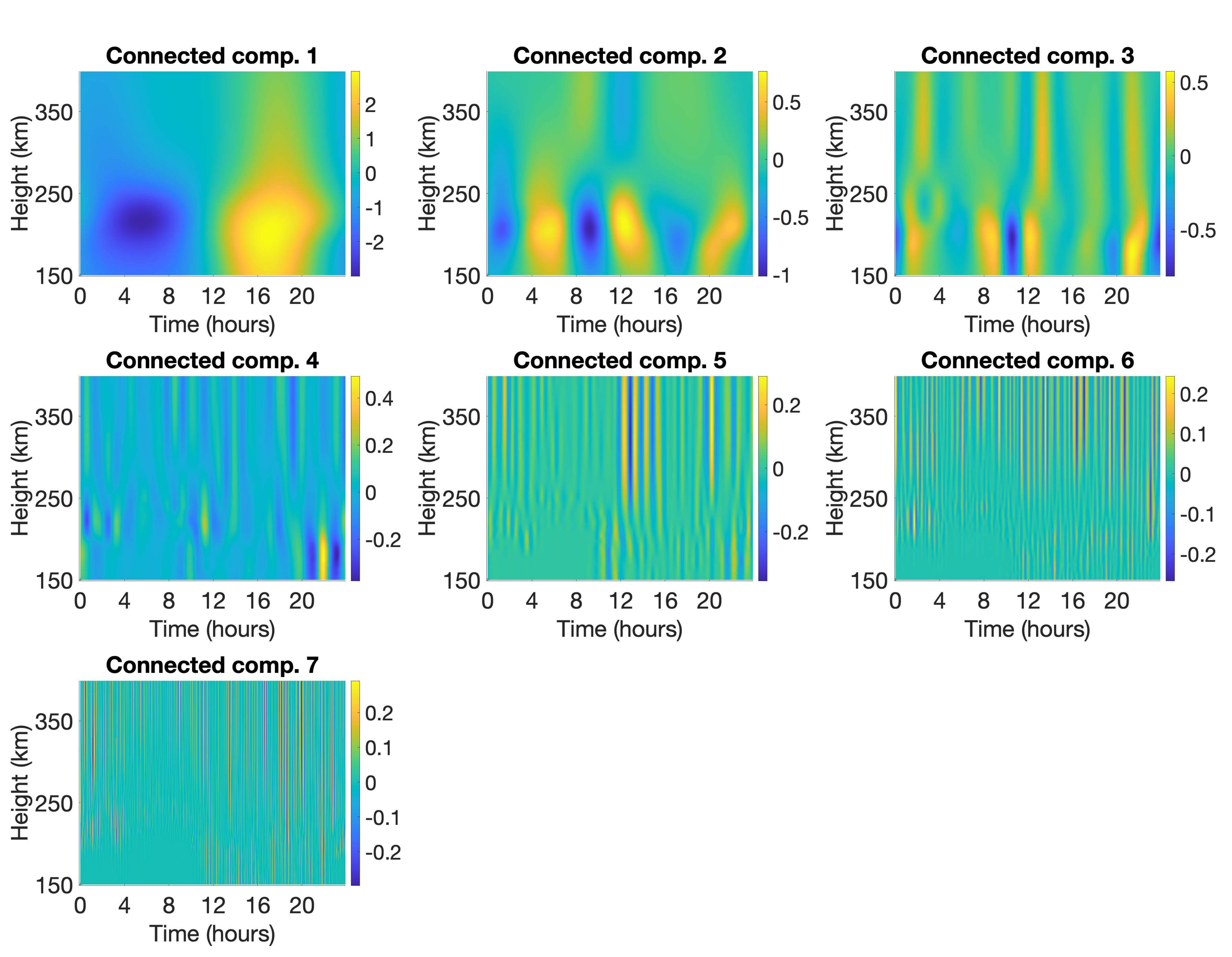}
    \caption{Dataset 1 connected components averaged over 24-hour lags, $\bar{\bf Y}_{n}^{C}$.}
    \label{fig:wave_group_cycles}
\end{figure}
Figure \ref{fig:wave_group_cycles} shows each $\bar{{\bf Y}}_{n}^{\text{C}}$ for Dataset 1. These matrices represent the average plasma frequency oscillation over a given day at various scales in the dynamics. Therefore, this step acts as a denoising process that has minimal impact on the multiscale nature of the signal and reduces the amount of information that would be lost by simply filtering the raw EDP time series.

Finally, using Equation \eqref{eqn:dmdmodel} on these 24-hour averaged and scale-correlated data, we generate a continuous-time DMD model for each connected component,
\begin{equation}
	\bar{{\bf y}}_{n}^{\text{C}}(t) \approx \bm{\Phi}_{n} \bm{\Lambda}_{n}^{t / \Delta t} \bm{\Phi}_{n}^{\dagger}\bar{{\bf y}}_{n,0}^{\text{C}}.
\end{equation}
Note that all of the $N_{C}$ components sum coherently and form the final the SSDMD model,
\begin{equation} \label{eqn:ssdmdmodel}
	\bar{{\bf y}}(t) \approx \sum_{n=1}^{N_{C}} {\bf W}_{n} \bm{\Lambda}_{n}^{t / \Delta t} {\bf W}_{n}^{\dagger}\bar{{\bf y}}_{n,0}^{\text{C}}.
\end{equation}

\begin{figure}[h!]
    \centering
    \includegraphics[width=1\textwidth]{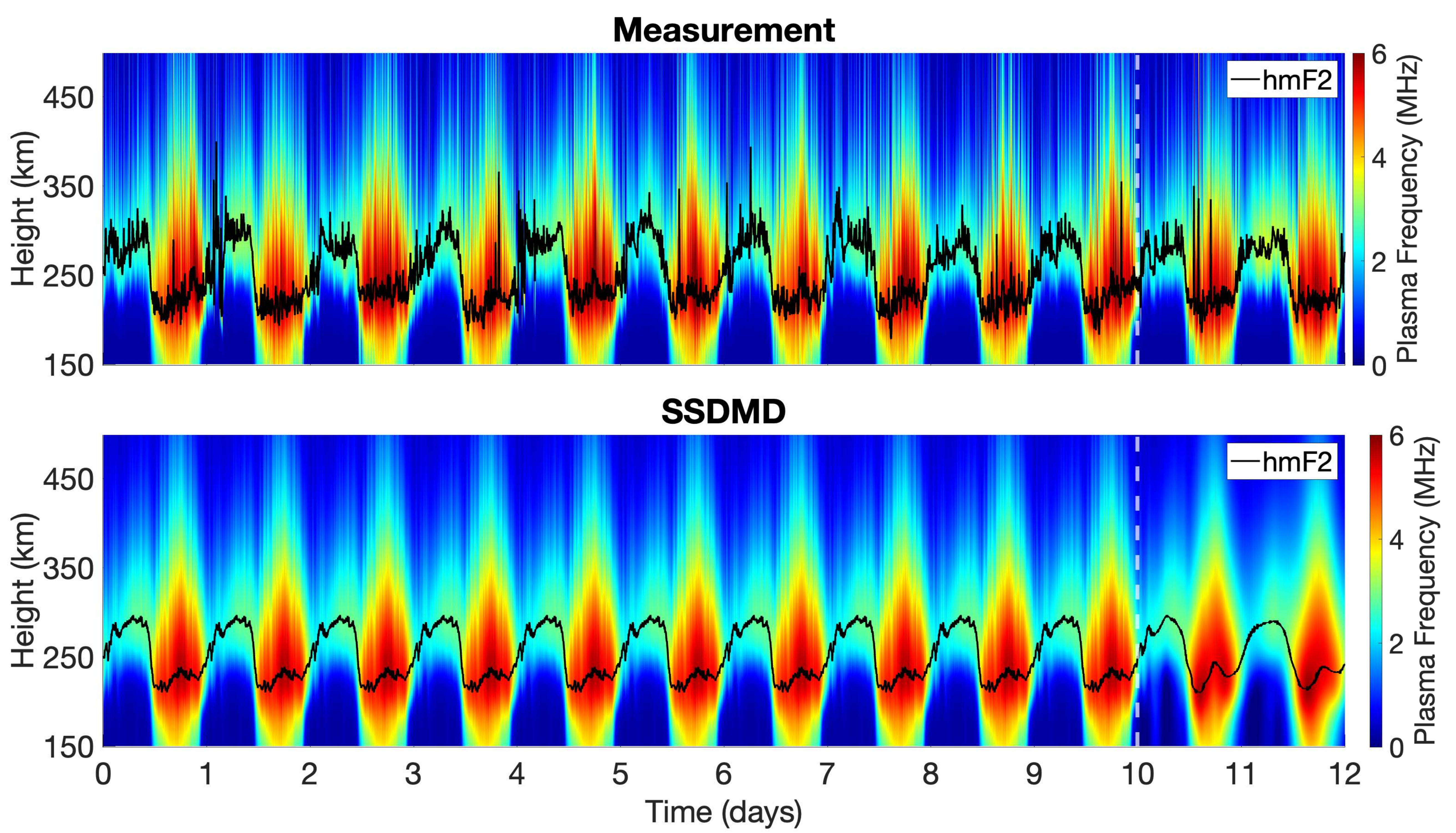}
    \caption{SSDMD reconstruction and forecast of Dataset 1. The vertical dotted white line denotes the transition from data used to fit the model to validation data. Black lines in each panel trace the hmF2 parameter.}
    \label{fig:recon_with_pred_bc840}
\end{figure}
Equation \eqref{eqn:ssdmdmodel} is a model for the dynamics of the average that accounts for nonlinear oscillations at multiple scales while preserving strong couplings between scales. See Appendix \ref{sec:algorithm} for pseudocode of the complete SSDMD algorithm. Figure \ref{fig:recon_with_pred_bc840} depicts the result of this model applied to Dataset 1, using the first 10 days of data to generate the SSDMD model and then advancing the DMD modes via their eigenvalues out an additional 2 days as a forecast. The figure includes both the original measurement time series and the SSDMD reconstruction and forecast.

We compute the foF2 and hmF2 parameters by finding the peak frequency and height in the modeled EDPs. Figure \ref{fig:recon_with_pred_bc840} shows the predicted hmF2 and observed hmF2 overlayed on their respective EDP time series. The reconstruction of the first 10 days, i.e. the fitting data, appears excellent simply because it is advancing each profile a single time step. The remaining two days, however, illustrate the stability of the modes that have been determined through SSDMD, since we are iterating the DMD eigenvalues and using the last observed EDP from the training data as an initial condition. Thus, we have built a stable time-stepping model of foF2 and hmF2 using a dynamical model that utilizes the full EDP time series expanded over several time scales. In Section \ref{sec:ssdmdresults} we will explore the accuracy of the resultant foF2 and hmF2 forecasts in greater detail.

\section{Results}\label{sec:results}
\subsection{Data Description}\label{sec:datadesc}
Data sets were gathered from Boulder, Colorado (40\textdegree N, -105.3\textdegree W) over 2019, and from Rome, Italy (41.9\textdegree N, 12.5\textdegree E) over 2014. The years 2019 and 2014 were roughly at the last solar minimum and solar maximum, respectively. These data sets will provide statistical estimates of how the proposed method performs at mid-latitudes during periods of high and low solar activity. Additionally, shorter data sets taken from Gakona, Alaska (62.38\textdegree N, 145\textdegree W) and Guam (13.62\textdegree N, 144.86\textdegree E) and will demonstrate the method's application in high-latitude and equatorial environments, respectively. Results presented for foF2 are in units of megahertz and hmF2 in kilometers unless otherwise labeled.

\begin{figure}[h!]
    \centering
    \includegraphics[width=1\textwidth]{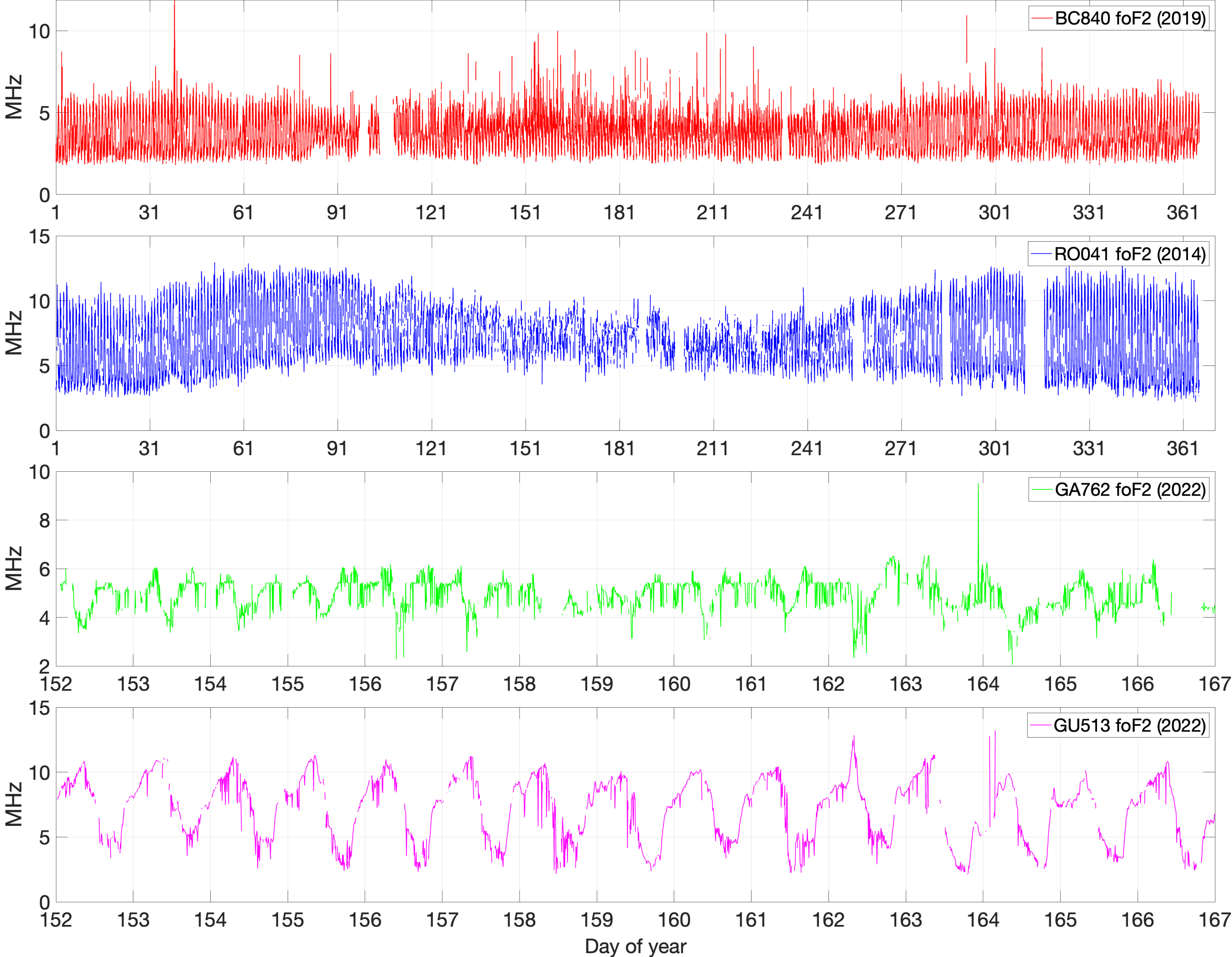}
    \caption{Time series of foF2 from the BC840 (red), RO041 (blue), GA762 (green), and GU513 (magenta) sounders. Note that the x-axis (day of year) has been zoomed in for the shorter data sets GA762 and GU513.}
    \label{fig:fof2_ts}
\end{figure}

\begin{figure}[h!]
    \centering
    \includegraphics[width=1\textwidth]{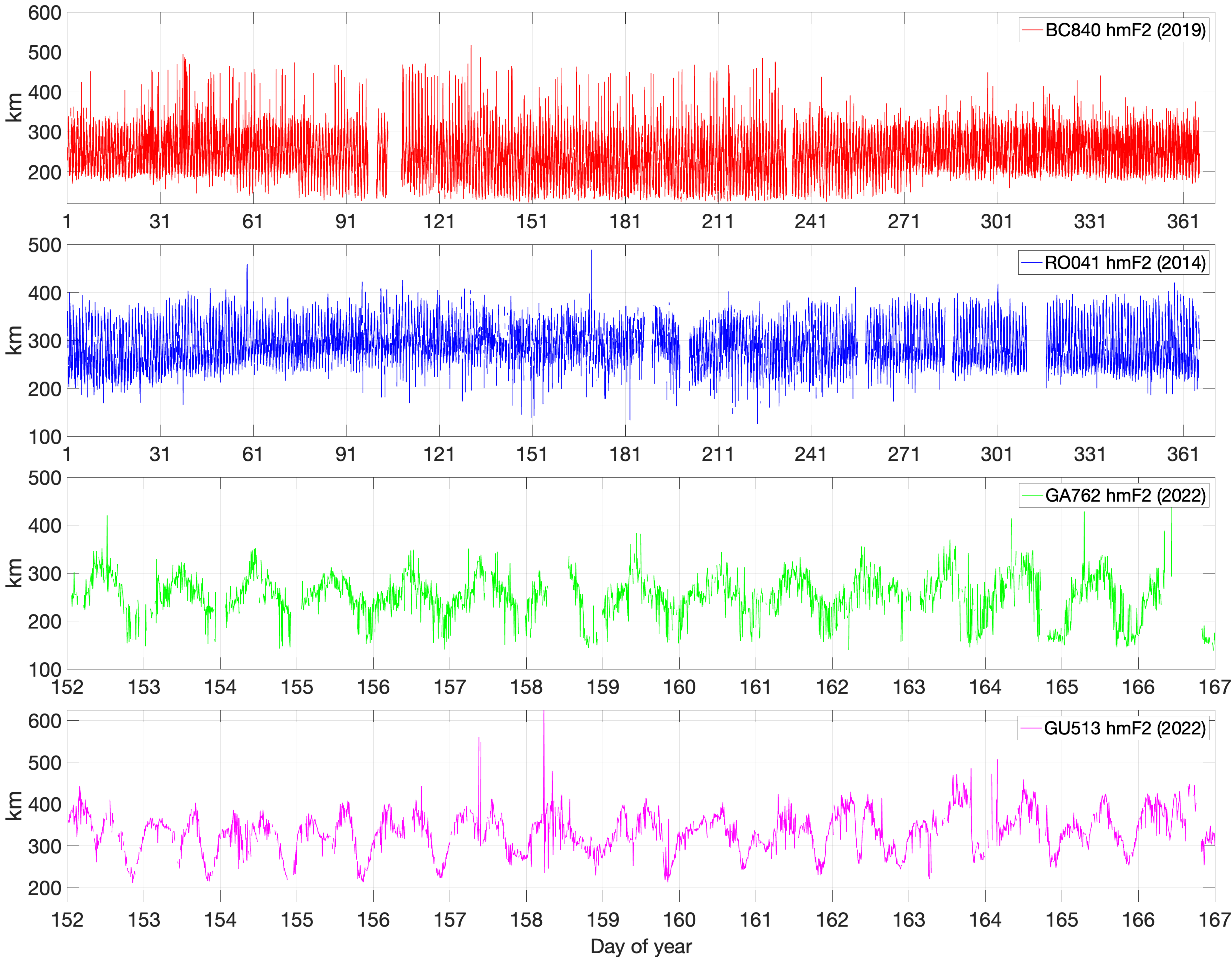}
    \caption{Time series of hmF2 from the BC840 (red), RO041 (blue), GA762 (green), and GU513 (magenta) sounders. Note that the x-axis (day of year) has been zoomed in for the shorter data sets GA762 and GU513.}
    \label{fig:hmf2_ts}
\end{figure}

The sounder located in Boulder, Colorado (station name BC840) had a measurement cadence of 5 minutes in 2019, while the Rome, Italy sounder (station name RO041) measured profiles every 15 minutes in 2014. The shorter data sets from Gakona, Alaska (station name GA762) and Guam (station name GU513) both had cadences of 7.5 minutes. Table \ref{tab:summary} summarizes the locations, times, and lengths of the data sets gathered for this study, and Figures \ref{fig:fof2_ts} and \ref{fig:hmf2_ts} show time series of the foF2 and hmF2 parameters as measured at each station. Each data point in these time series has an affiliated EDP, but these are not shown for brevity. Missing values in the data are not used in the final error analysis.

All sounder stations generate estimates of the vertical EDP using the ARTIST5 algorithm to invert raw ionograms \cite{galkin_2008}. The EDP time series is limited to a height range of 150-500km. This is primarily because the plasma frequency in E-region at night dips low enough that it is outside the measurement bandwidth of the Digisonde sounders \cite{Bibl_1981}. Because of this, the ARTIST5 inversion algorithm will generally output a default value, e.g., 0.2 MHz, in these regions for most of the nighttime profiles. These periods of constant plasma density complicate the fitting of an SSDMD model since they require inherently oscillatory modes to approximate a constant value. Above the peak plasma density, echoes from the sounder are no longer received, and a standard parameterized profile is fit to provide the topside plasma density. Thus, restricting the profiles to only the F-region helps ensure the SSDMD model is able to more accurately capture the dynamics of the F-layer parameters and minimizes the effects of these boundary regions.
\begin{table}[h!]
    \centering
    \begin{tabular}{l c c c c c}
    \hline
    \textbf{} & \textbf{\begin{tabular}[c]{@{}c@{}}Boulder\end{tabular}} & \textbf{\begin{tabular}[c]{@{}c@{}}Rome\end{tabular}} & \textbf{\begin{tabular}[c]{@{}c@{}}Gakona\end{tabular}} & \textbf{\begin{tabular}[c]{@{}c@{}}Guam \end{tabular}}\\
    \hline
        Station name & BC840 & RO041 & GA762 & GU513 \\ \hline
        Year  & 2019 & 2014 & 2022 & 2022 \\ \hline
        Lat/Lon & 40\textdegree N 105.3\textdegree W & 41.9\textdegree N 12.5\textdegree E & 62.38\textdegree N 145\textdegree W & 13.62\textdegree N 144.86\textdegree E\\ \hline
        Number of days  &  365  &  365 & 12 & 12  \\ \hline
        Measurement cadence  &  5 min.  &  15 min.  &  7.5 min.  &  7.5 min. \\  \hline
        Solar cycle & min & max & mid & mid \\
    \hline
    \end{tabular}
    \caption{Summary of data gathered from Didbase sounder stations.}
    \label{tab:summary}
\end{table}

We used the IRI2016 model in Python with up-to-date solar and magnetic indices. IRI has many settings that allow the user to tweak parameters or turn certain submodels on or off. These settings are known as the {\it JF} switches. The version of IRI used in this paper had all the default {\it JF} values, which are found on the IRI model website. Time series of the EDP, foF2, and hmF2 were generated from IRI for each data set, and the EDPs were interpolated to the same vertical height grid as the sounder data.

There are several hyperparameters of the SSDMD model that must be set prior to fitting a model. The first is the correlation threshold from Equation \ref{eqn:corr_thresh} that determines how strongly scales must be correlated in order to form a connected component. This threshold currently requires manual tuning. We found a value of $c_{corr}=-1.95$ achieved good results for stations BC840, GA762, and GU513, while $c_{corr}=-1.75$ performed better for RO041. Generating more efficient ways of determining the optimal value for this parameter will be a topic of future research, though its value here was chosen such that the MAE of the foF2 and hmF2 parameters were minimized. 

Another hyperparameter is the number of days used to fit the SSDMD model. Using long time series will result in more averaging over the 24-hour cycles, thus increasing bias in the forecast. We found that 10 days of EDPs worked reasonably well for all stations for short-term prediction. If one attempts a longer-term forecast, averaging over additional time lags may be necessary. The last hyperparameter of SSDMD is the threshold at which to truncate the singular values in the DMD step, Equation \ref{eqn:dmd_thresh}. This threshold was set to $c_{svd}=6$, which worked well for all data sets. Lowering this threshold will result in fewer spectral pairs $(\lambda_{j}, {\bf w}_{j})$ in the SSDMD model and thus reduces the number of modes used to generate the forecast. Table \ref{tab:params_summary} summarizes these hyperparameters.
\begin{table}[h!]
    \centering
    \begin{tabular}{l c c}
    \hline
    \textbf{\begin{tabular}[c]{@{}c@{}}SSDMD Parameter\end{tabular}} & \textbf{\begin{tabular}[c]{@{}c@{}}Value\end{tabular}}\\
    \hline
        Num. days for fit & 10 \\ \hline
        Num. days forecast  & 2 \\ \hline
        $c_{corr}$ & -1.95 (BC840, GA762, GU513) / -1.75 (RO041)\\ \hline
        $c_{svd}$  &  6 \\ \hline
        Wavelet type  &  coiflet $4^{th}$ order \\
    \hline
    \end{tabular}
    \caption{Summary of parameters for the SSDMD model used for each data set.}
    \label{tab:params_summary}
\end{table}

\subsection{SSDMD Model Performance}\label{sec:ssdmdresults}
We tested the SSDMD method on 30 randomly chosen 12-day periods in the BC840 and RO041 data sets. Each of these stations contained several large gaps in their data which were not used in the random start times as one cannot fit an SSDMD model without contiguous data. Even though standard DMD methods will work for arbitrary snapshots of data $({\bf x}, {\bf y})$, where ${\bf y}={\bf K}{\bf x}$, the wavelet decompositions used in SSDMD require a regular measurement cadence, i.e., the data snapshots are always $\delta t$ time apart. 

\begin{figure}[t!]
    \centering
    \includegraphics[width=1\textwidth]{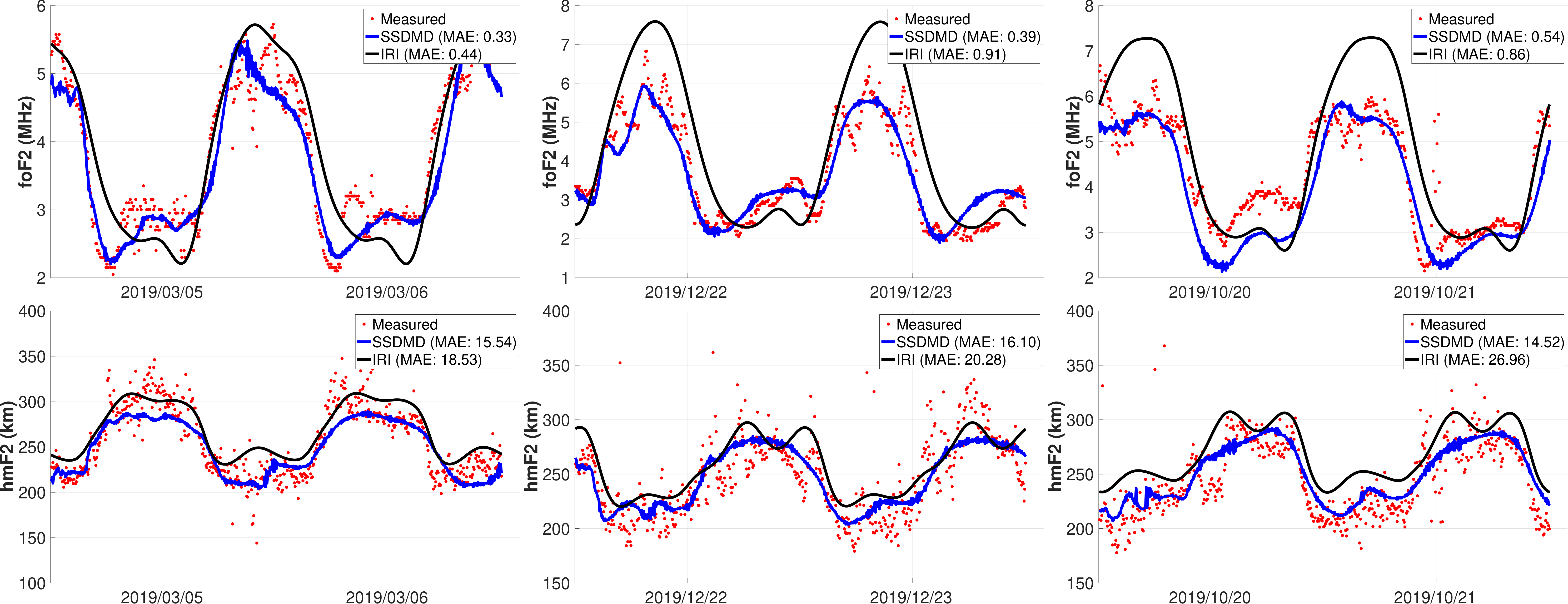}
    \caption{SSDMD forecasts of foF2 (top panels) and hmF2 (bottom panels) for the BC840 sounding station for randomly chosen starting times in 2019. The MAE is provided for both the SSDMD and IRI forecasts.}
    \label{fig:forecasts_bc840}
\end{figure}

\begin{figure}[t!]
    \centering
    \includegraphics[width=1\textwidth]{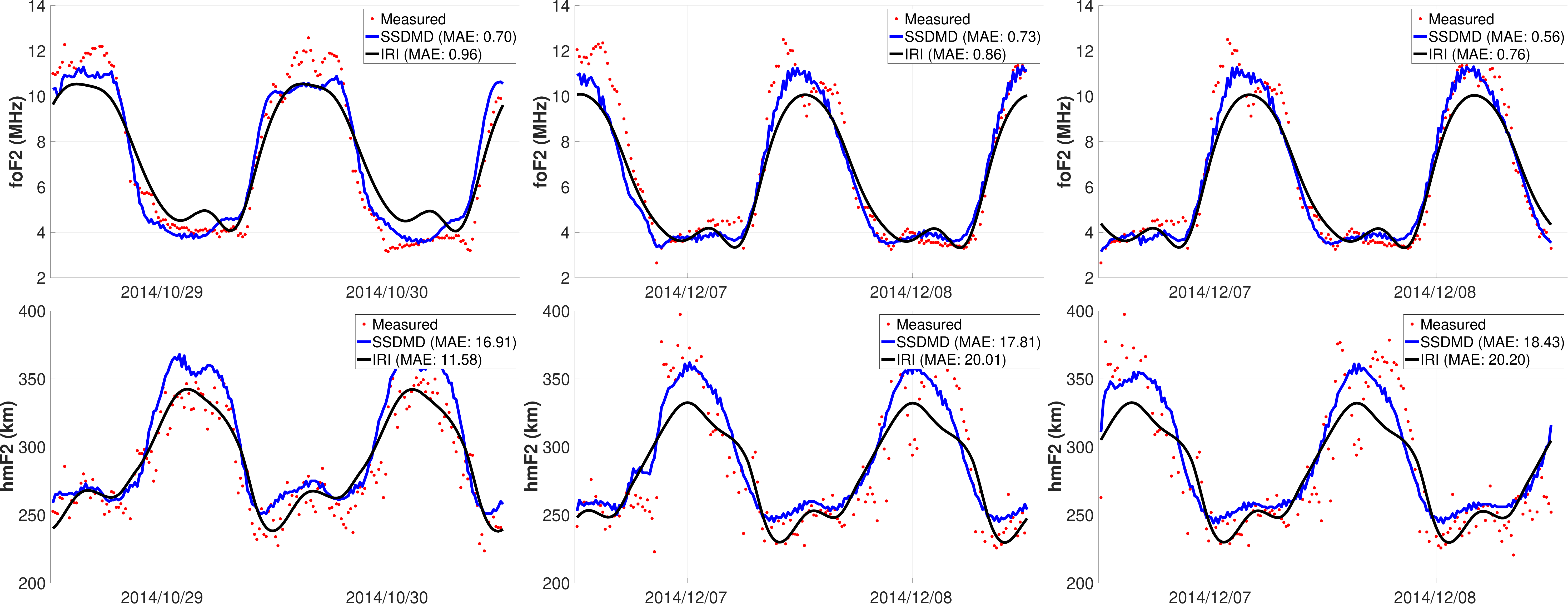}
    \caption{SSDMD forecasts of foF2 (top panels) and hmF2 (bottom panels) for the RO041 sounding station for randomly chosen starting times in 2014. The MAE is provided for both the SSDMD and IRI forecasts.}
    \label{fig:forecasts_ro041}
\end{figure}

For each random 12-day period, the first 10 days were used for fitting an SSDMD model and the remaining 2 days for testing a 48-hour forecast of the foF2 and hmF2 parameters. Figures \ref{fig:forecasts_bc840} and \ref{fig:forecasts_ro041} show these test forecast periods for 3 of the 30 randomly chosen times in each of the BC840 and RO041 data sets. The SSDMD and IRI predictions for the F-layer parameters, along with the measured values from the sounder, are presented for each. From these, we see that SSDMD captures some smaller-scale fluctuations in the parameters that are commonly lost in climatological models due to extreme averaging over monthly and seasonal variations. The mean absolute error (MAE) is provided for each forecast. While, in general, the SSDMD MAE shows modest improvements over IRI for BC840 in 2019, it is not always the case, as we can see in the hmF2 forecast for RO041 in 2014. However, in the cases where SSDMD does perform worse than IRI, it is still relatively close considering how little data is used to generate the forecast.

\begin{figure}[t!]
    \centering
    \includegraphics[width=1\textwidth]{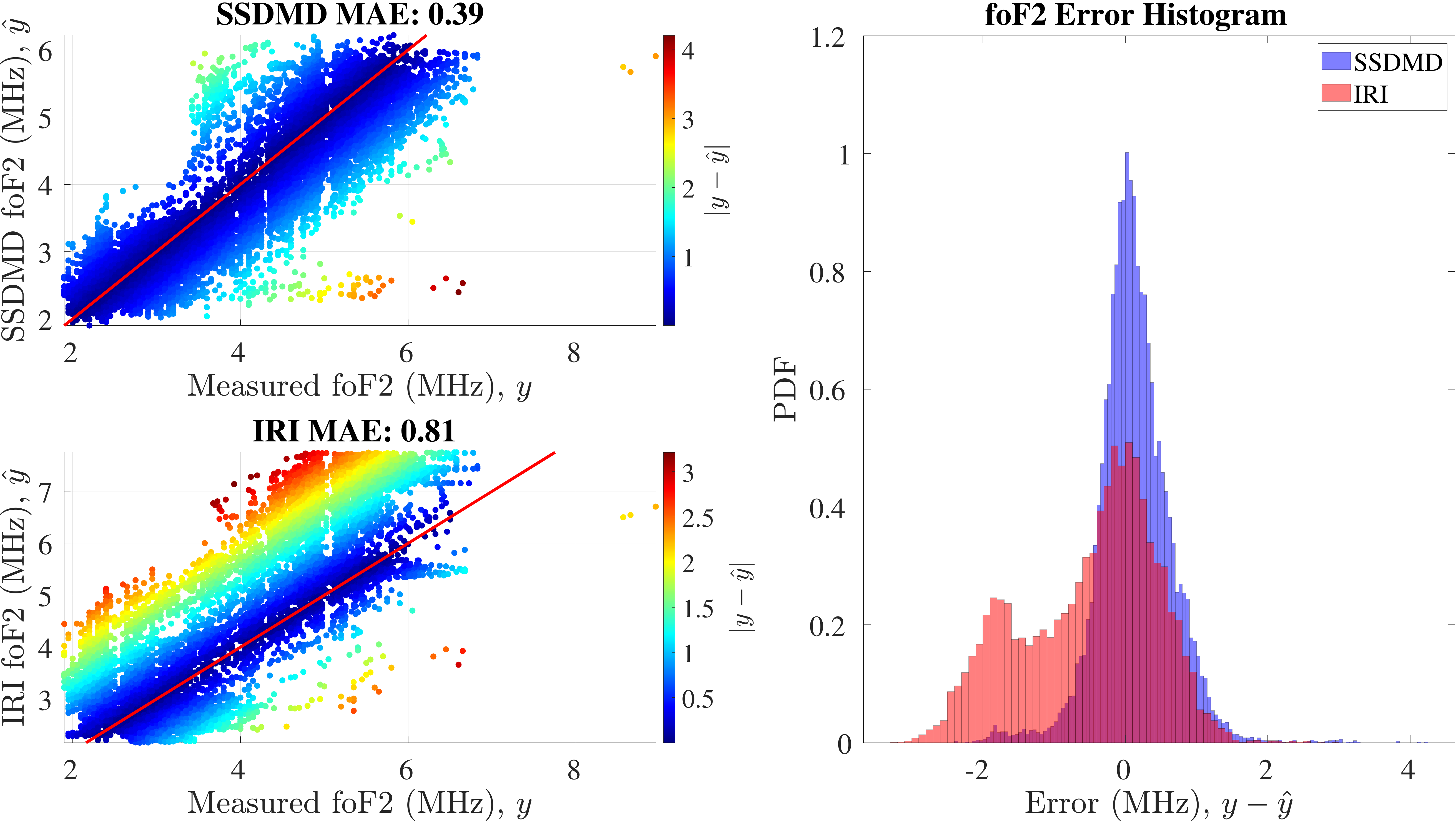}
    \caption{Forecasted vs. measured foF2 parameter scatter plots for the SSDMD (top left) and IRI (bottom left) models for the BC840 station in 2019. The total MAE for each model is given above their respective scatter plot. Histograms (right) provide estimates of the total model error distributions.}
    \label{fig:fof2_scatter_bc840}
\end{figure}
\begin{figure}[t!]
    \centering
    \includegraphics[width=1\textwidth]{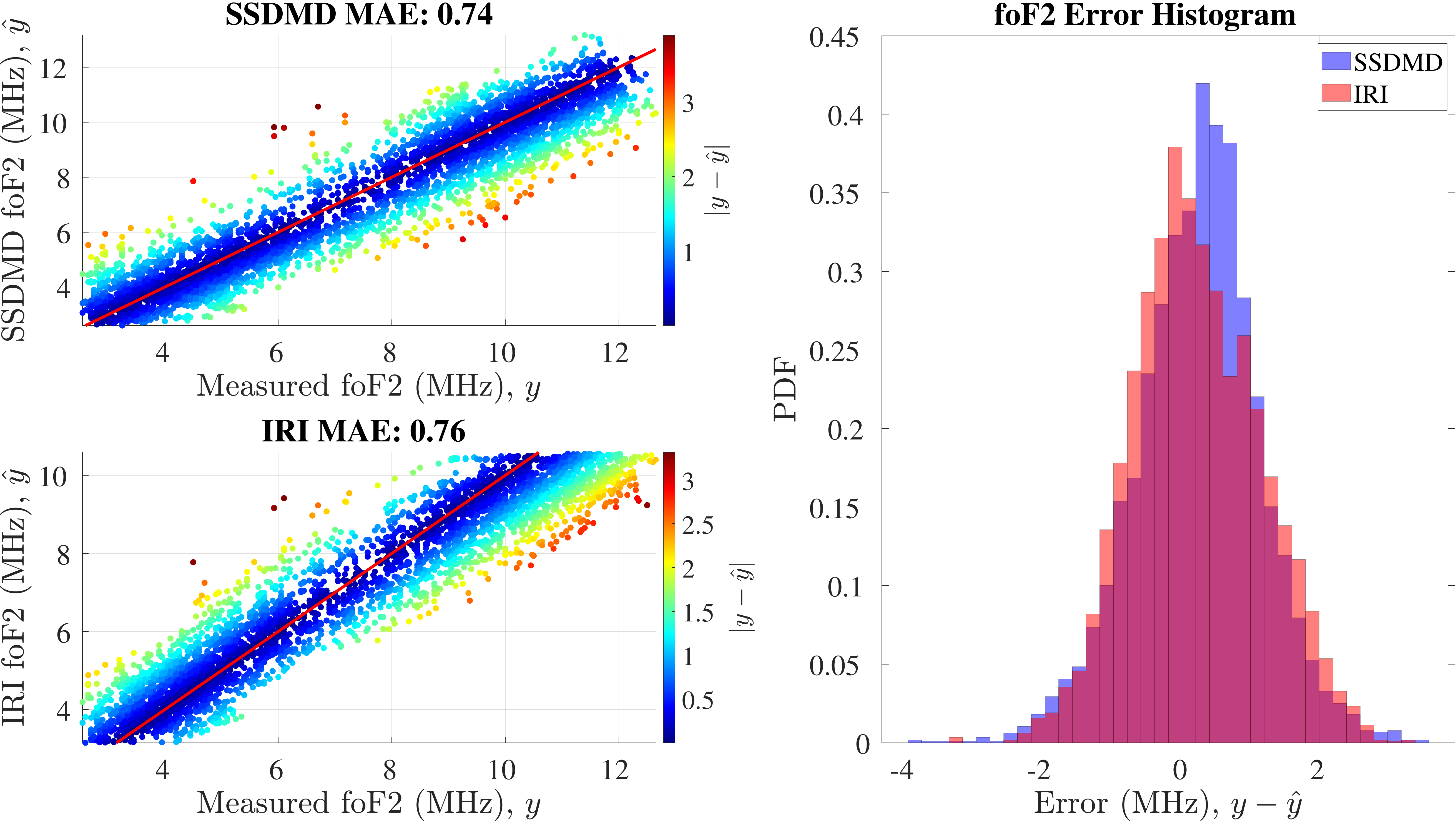}
    \caption{Forecasted vs. measured foF2 parameter scatter plots for the SSDMD (top left) and IRI (bottom left) models for the RO041 station in 2014. The total MAE for each model is given above their respective scatter plot. Histograms (right) provide estimates of the total model error distributions.}
    \label{fig:fof2_scatter_ro041}
\end{figure}

Figures \ref{fig:fof2_scatter_bc840} and \ref{fig:fof2_scatter_ro041} provide scatter plots and histograms of the foF2 modeled vs. measured forecasts for BC840 and RO041, respectively. The histograms are given to illustrate the shapes of the total model error distributions. The area of each bin simply represents the relative number of model errors within that interval over all 48-hour forecast test periods. Note that SSDMD forecasts perform markedly better on the BC840 data set, with IRI producing a significant bimodal error distribution. This may point toward limitations in SSDMD's applicability during periods of high solar activity. Figure \ref{fig:fof2_ts} shows a significant seasonal variation in the foF2 parameter of the RO041 station. Applying SSDMD to longer time series to capture seasonal and solar cycle trends will be a topic of future study. Furthermore, in the context of short-term forecasts, SSDMD's reliance on the fit of the ${\bf K}_o$ matrix to advance any data point one time-step into the future benefits from higher measurement cadences. In addition, as the time resolution of sounder measurements increases, a wider spectrum of geophysical noise will be observed, and thus, SSDMD's ability to identify couplings between dominant scales becomes more pronounced.

\begin{figure}[h!]
    \centering
    \includegraphics[width=1\textwidth]{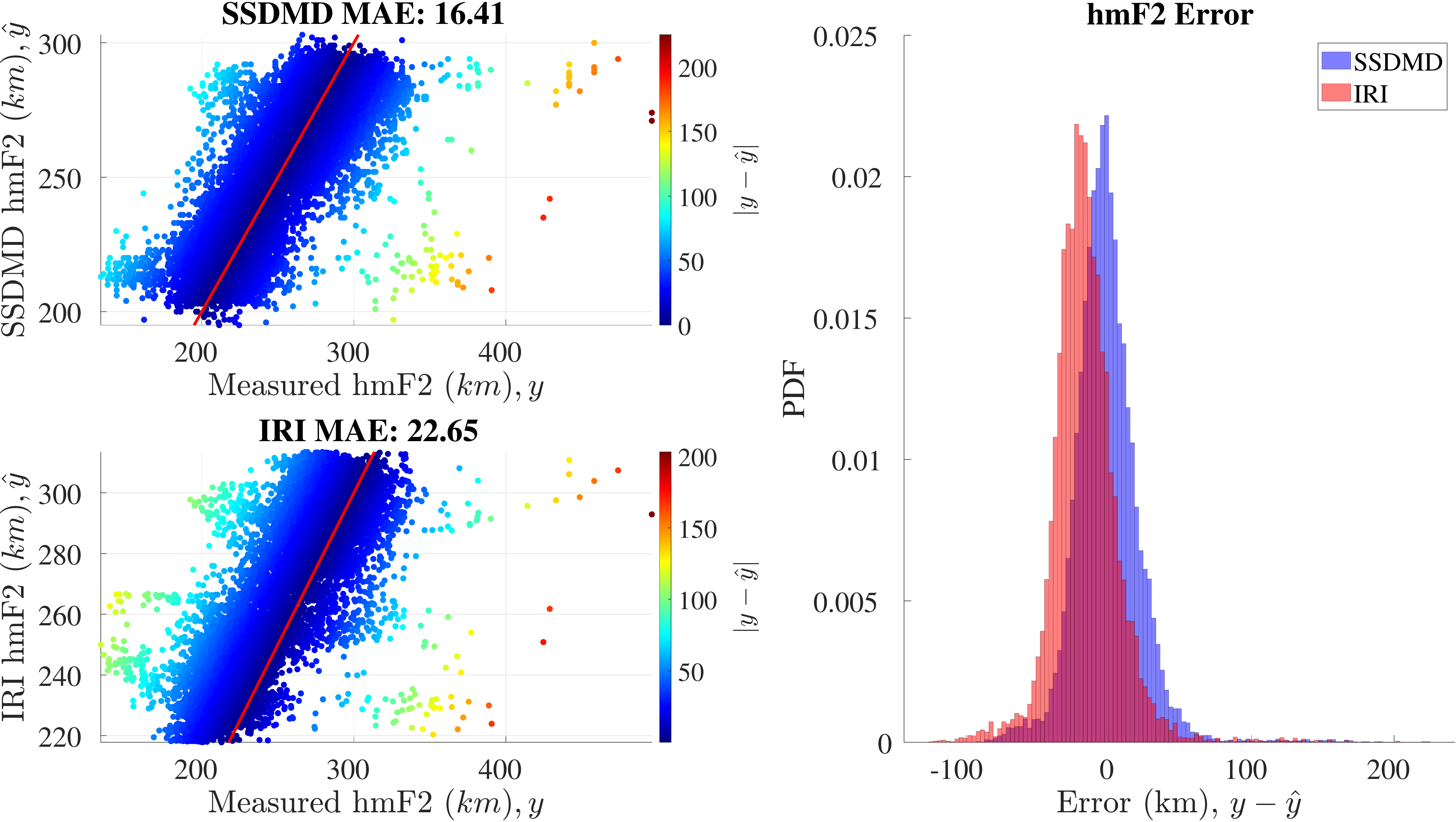}
    \caption{Forecasted vs. measured hmF2 parameter scatter plots for the SSDMD (top left) and IRI (bottom left) models for the BC840 station in 2019. The total MAE for each model is given above their respective scatter plot. Histograms (right) provide estimates of the total model error distributions.}
    \label{fig:hmf2_scatter_bc840}
\end{figure}
\begin{figure}[h!]
    \centering
    \includegraphics[width=1\textwidth]{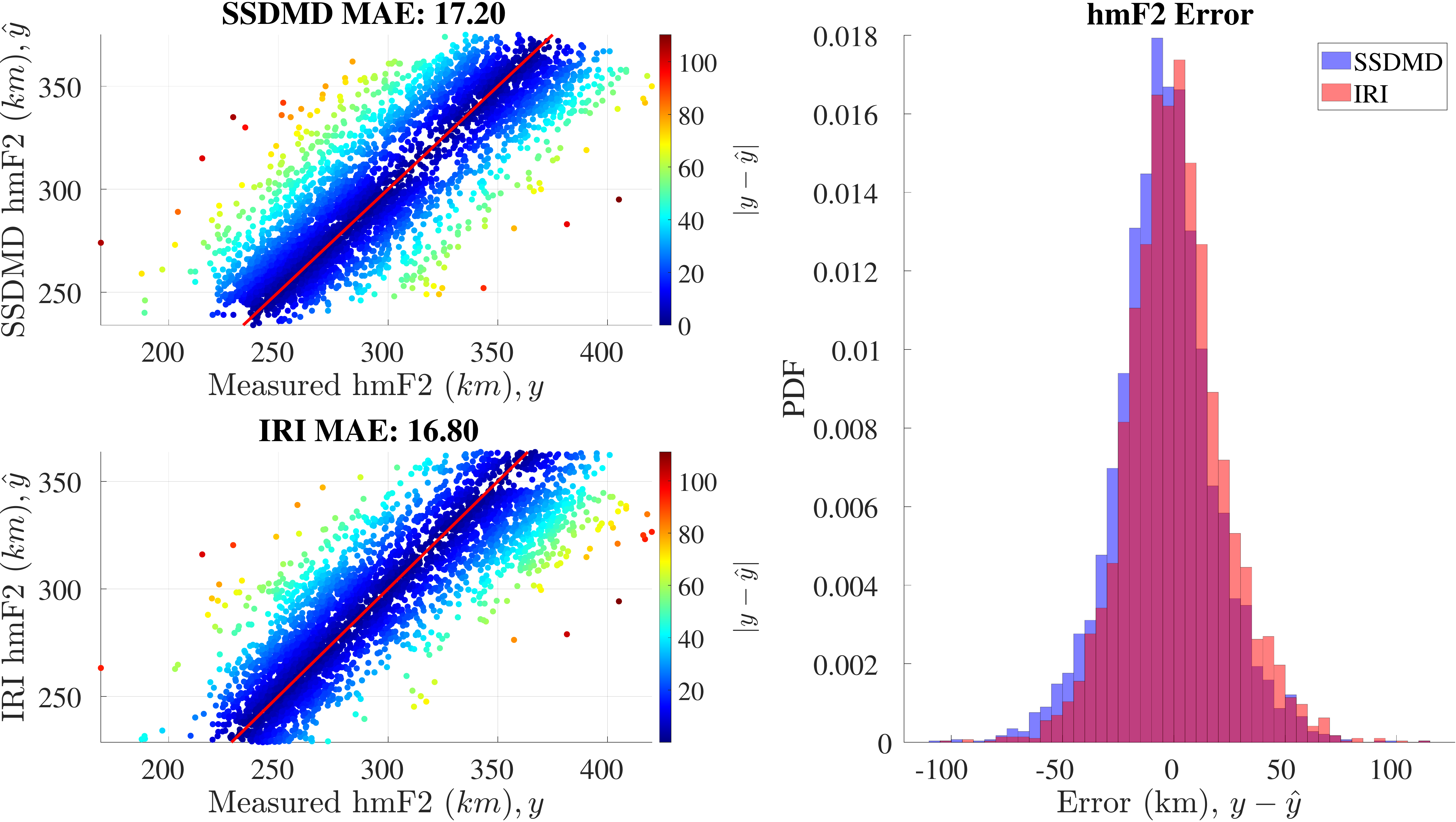}
    \caption{Forecasted vs. measured hmF2 parameter scatter plots for the SSDMD (top left) and IRI (bottom left) models for the RO041 station in 2014. The total MAE for each model is given above their respective scatter plot. Histograms (right) provide estimates of the total model error distributions.}
    \label{fig:hmf2_scatter_ro041}
\end{figure}
Figures \ref{fig:hmf2_scatter_bc840} and \ref{fig:hmf2_scatter_ro041} give similar scatter plots and histograms for the hmF2 parameter for the BC840 and RO041 stations, respectively. With hmF2, we find the model error distributions for both SSDMD and standard IRI to be very similar. However, SSDMD provides a slight bias correction over IRI for the BC840 data set. While the hmF2 MAE for SSDMD on the RO041 data is worse than IRI, its performance is still quite close, given the relatively small amount of data used to generate the forecast.

\begin{figure}[h!]
    \centering
    \includegraphics[width=0.55\textwidth]{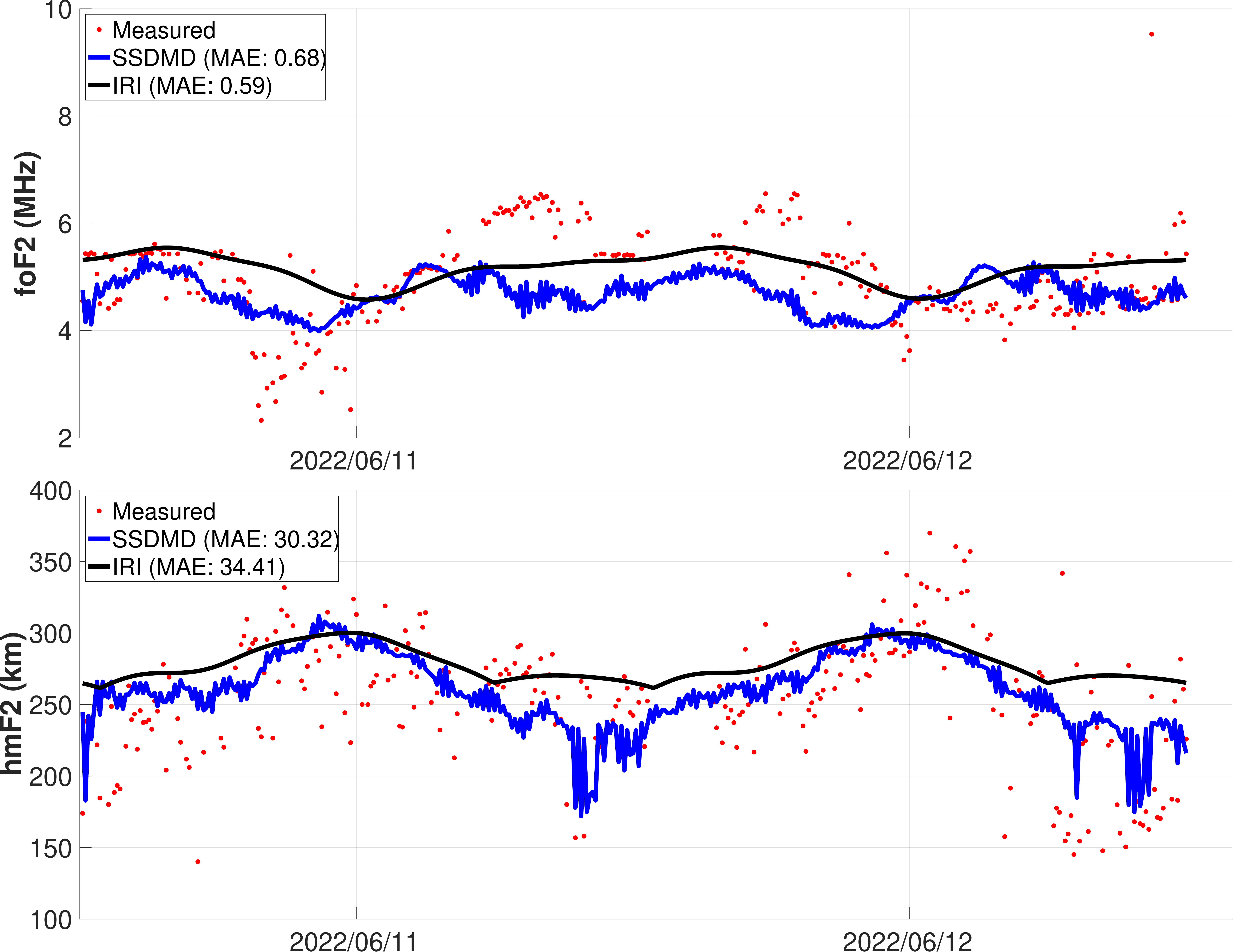}
    \caption{SSDMD 2-day forecast of the foF2 (top) and hmF2 (bottom) parameters for the GA762 station with IRI predictions. MAE values for both models are provided in the legend.}
    \label{fig:forecasts_ga762}
\end{figure}
\begin{figure}[h!]
    \centering
    \includegraphics[width=0.75\textwidth]{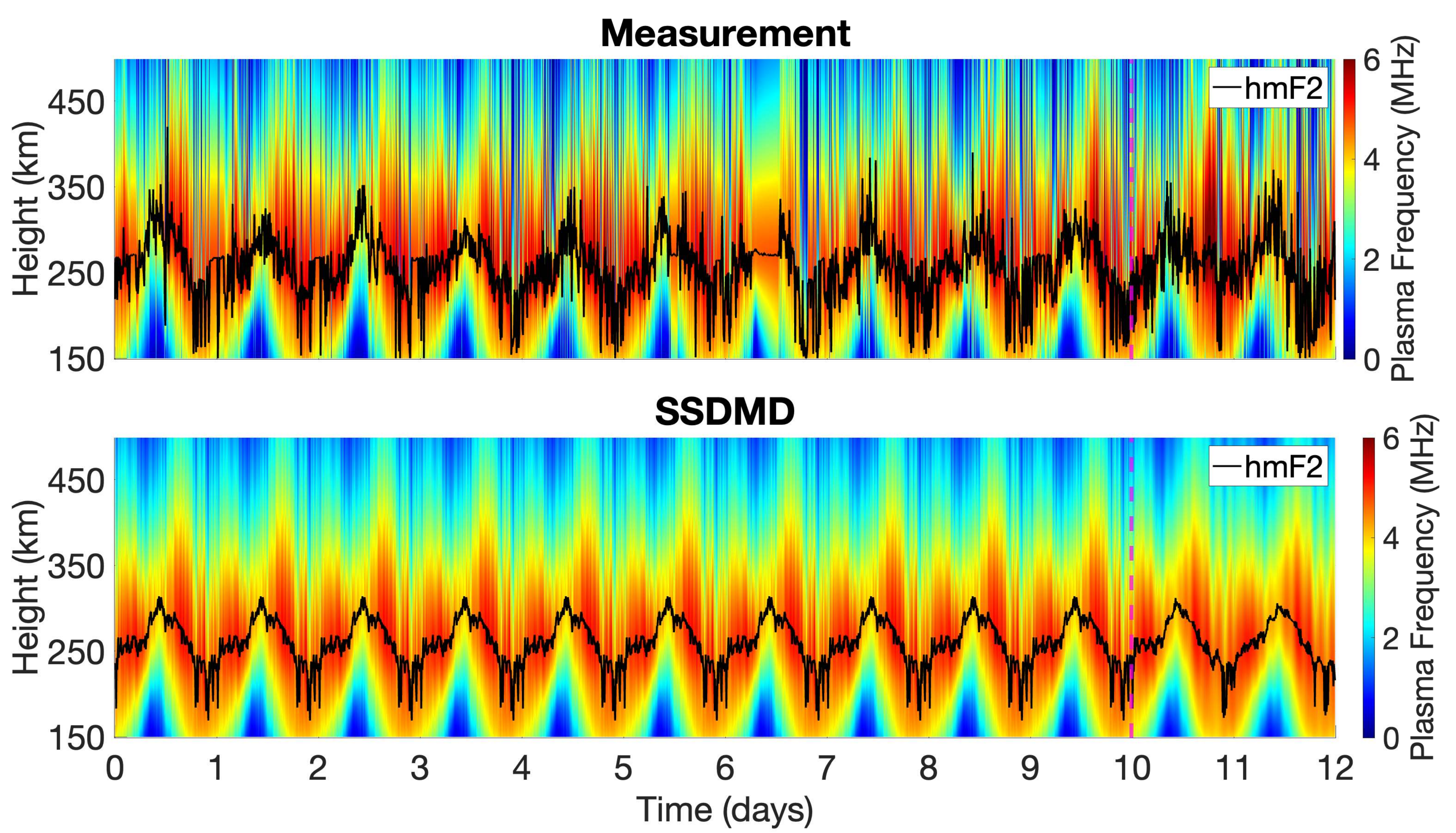}
    \caption{SSDMD full EDP time series reconstruction and 2-day forecast of the GA762 station during a 15-day period in 2022. The vertical dotted magenta line indicates the transition from fitting data to test data, and the solid black line follows the hmF2 parameter computed using the EDP time series.}
    \label{fig:recon_with_pred_ga762}
\end{figure}
The SSDMD model was run on the GA762 station data set to illustrate its use on data streams from higher latitudes. GA762 is at a latitude of 62.38\textdegree N and is the site of the High-frequency Active Auroral Research Program (HAARP) \cite{bailey_2000}, a valuable ionospheric-thermospheric research instrument used in a variety of fundamental and experimental physics applications \cite{bell_2001, bernhardt_2009}. Improved forecasts of the foF2 and hmF2 parameters continue to play a critical role in high-frequency radio experimentation and modeling. The use of a lightweight and adaptive forecast like SSDMD for real-time operations may be explored in future work, but in this paper we use this station to provide validation of our method in these high-latitude regions. Figures \ref{fig:forecasts_ga762} and \ref{fig:recon_with_pred_ga762} give forecasts of foF2 and hmF2 and visualizations of the full EDP reconstructions for this station.

\begin{figure}[h!]
    \centering
    \includegraphics[width=0.55\textwidth]{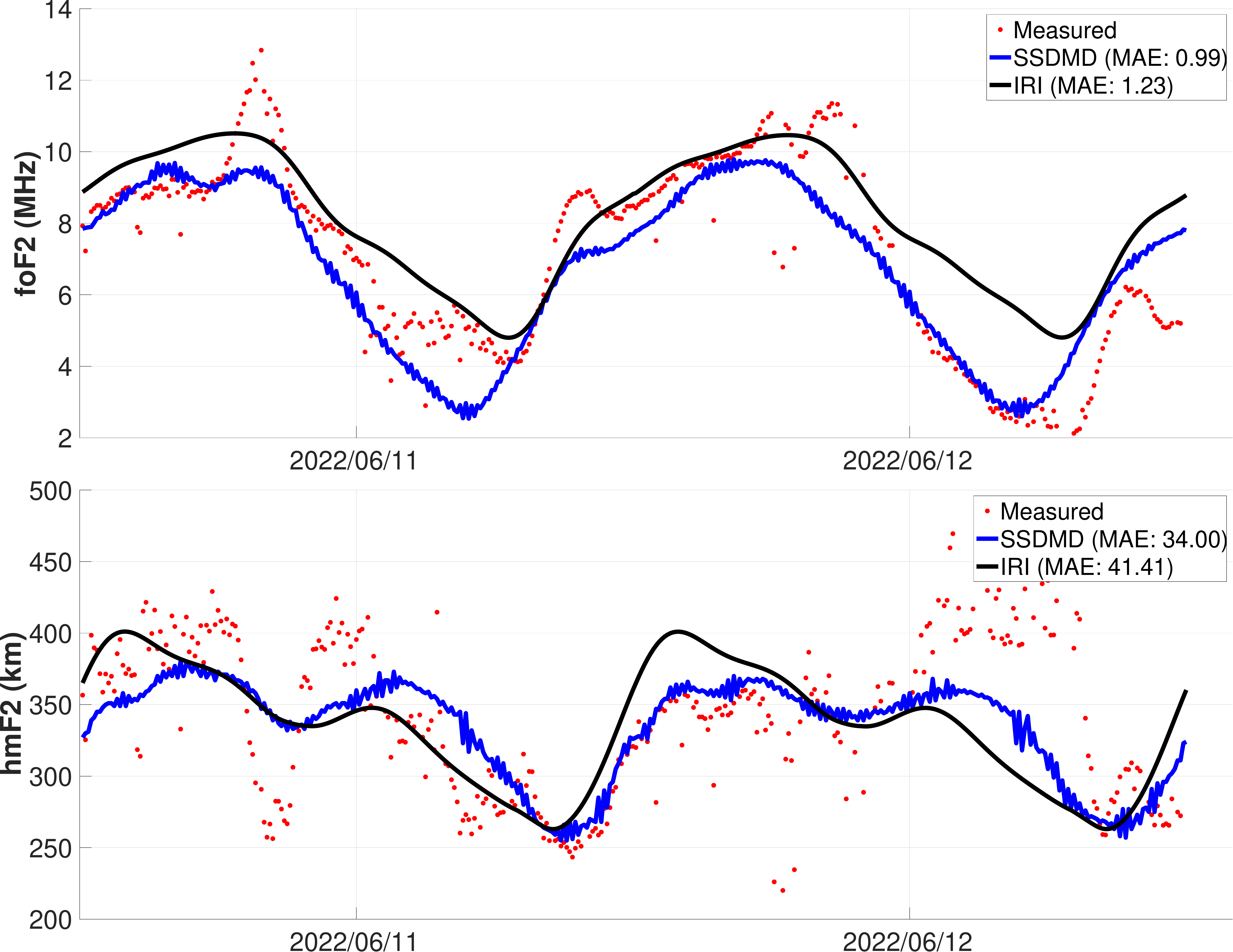}
    \caption{SSDMD 2-day forecast of the foF2 (top) and hmF2 (bottom) parameters for the GU513 station with IRI predictions. MAE values for both models are provided in the legend.}
    \label{fig:forecasts_gu513}
\end{figure}
\begin{figure}[h!]
    \centering
    \includegraphics[width=0.75\textwidth]{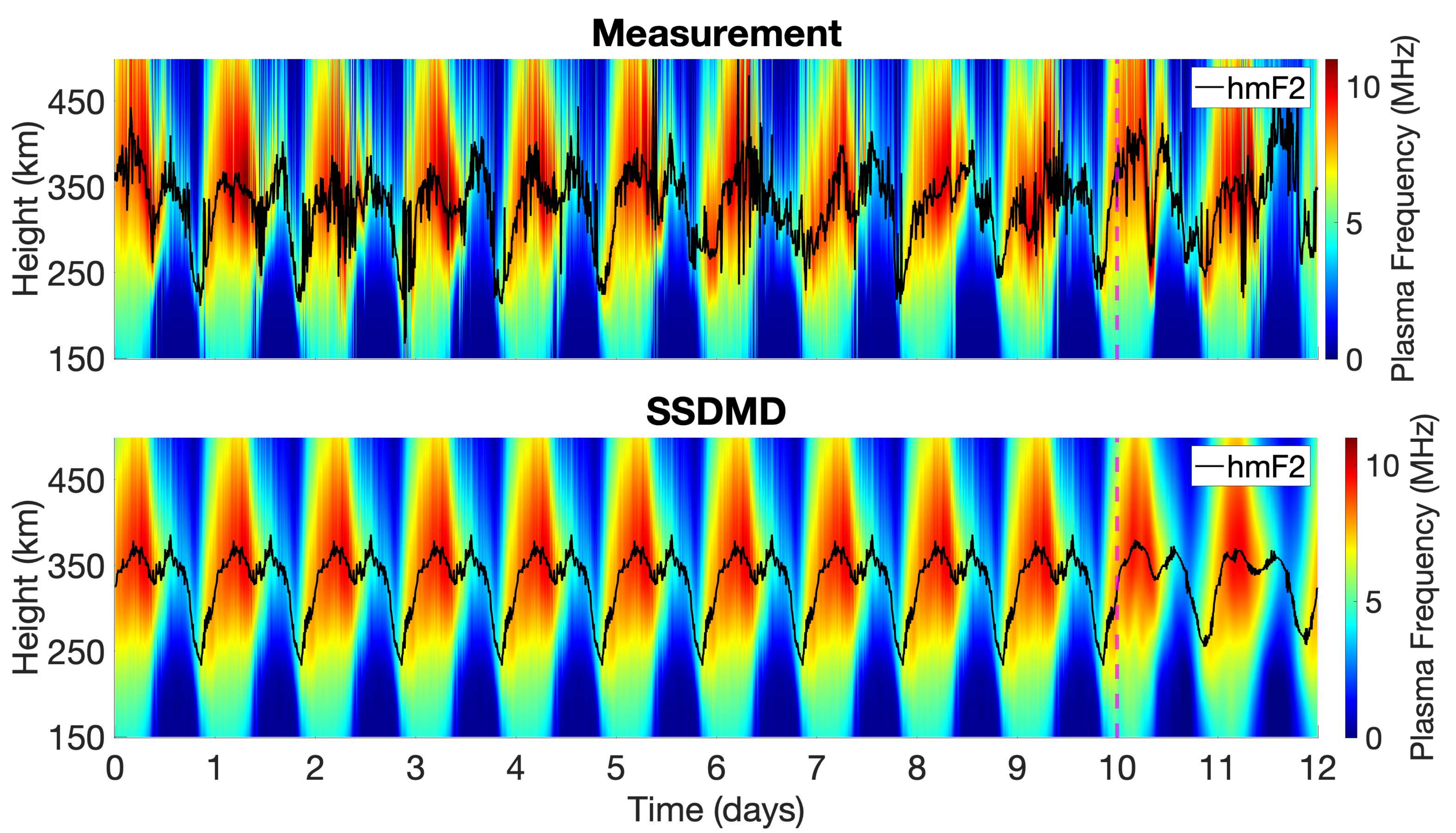}
    \caption{SSDMD full EDP time series reconstruction and 2-day forecast of the GU513 station during a 15-day period in 2022. The vertical dotted magenta line indicates the transition from fitting data to test data and the solid black line follows the hmF2 parameter computed using the EDP time series.}
    \label{fig:recon_with_pred_gu513}
\end{figure}
Lastly, Figures \ref{fig:forecasts_gu513} and \ref{fig:recon_with_pred_gu513} demonstrate the SSDMD model in a low-latitude environment. Figure \ref{fig:recon_with_pred_gu513} illustrates the dramatic oscillations of the hmF2 as compared with the mid- and high-latitude stations. The presence of complex physical processes like the equatorial plasma fountain \cite{macdougall_1969, balan_2018} induce categorically more complex dynamics in the EDP time series than observed at mid-latitudes. Still, we find SSDMD can fit a model that improves the MAE for both foF2 and hmF2 compared to IRI. 

In addition to the MAE statistics presented for each station, Tables \ref{tab:fof2_error} and \ref{tab:hmf2_error} give summaries of root-mean-squared error (RMSE) and mean absolute percentage error (MAPE) for all foF2 and hmF2 forecasts, respectively. We find that SSDMD either outperforms or closely matches a standard IRI forecast for both foF2 and hmF2 for the data sets presented. While significant improvement in the IRI forecast can be made by tweaking coefficients within the model or even through the assimilation of real time data, SSDMD provides an easily implementable fitting method that can adapt to new data in real-time. Moreover, adjusting the parameters within IRI will not always improve its forecast accuracy, as one does not know in which direction to adjust parameters until observations of the ionosphere are made.

\begin{table}[h!]
    \centering
    \begin{tabular}{ccccccc}
        \hline
        \multicolumn{7}{c}{\textbf{foF2 Forecast Errors}} \\ \hline
        \multicolumn{1}{c}{} & \multicolumn{2}{c}{RMSE} & \multicolumn{2}{c}{MAE} & \multicolumn{2}{c}{MAPE} \\ \hline
        \multicolumn{1}{c}{Station} & \multicolumn{1}{c}{SSDMD} & \multicolumn{1}{c}{IRI} & \multicolumn{1}{c}{SSDMD} & \multicolumn{1}{c}{IRI} & \multicolumn{1}{c}{SSDMD} & IRI \\ \hline
        \multicolumn{1}{c}{BC840} & \multicolumn{1}{c}{\textbf{0.54}} & \multicolumn{1}{c}{1.06} & \multicolumn{1}{c}{\textbf{0.39}} & \multicolumn{1}{c}{0.81} & \multicolumn{1}{c}{\textbf{10.08}} & 21.54 \\ \hline
        \multicolumn{1}{c}{} & \multicolumn{1}{c}{} & \multicolumn{1}{c}{} & \multicolumn{1}{c}{} & \multicolumn{1}{c}{} & \multicolumn{1}{c}{} &  \\ \hline
        \multicolumn{1}{c}{RO041} & \multicolumn{1}{c}{\textbf{0.93}} & \multicolumn{1}{c}{0.95} & \multicolumn{1}{c}{\textbf{0.74}} & \multicolumn{1}{c}{0.76} & \multicolumn{1}{c}{11.54} & \textbf{11.44} \\ \hline
        \multicolumn{1}{c}{} & \multicolumn{1}{c}{} & \multicolumn{1}{c}{} & \multicolumn{1}{c}{} & \multicolumn{1}{c}{} & \multicolumn{1}{c}{} &  \\ \hline
        \multicolumn{1}{c}{GA762} & \multicolumn{1}{c}{0.91} & \multicolumn{1}{c}{\textbf{0.81}} & \multicolumn{1}{c}{0.68} & \multicolumn{1}{c}{\textbf{0.59}} & \multicolumn{1}{c}{13.59} & \textbf{13.18} \\ \hline
        \multicolumn{1}{c}{} & \multicolumn{1}{c}{} & \multicolumn{1}{c}{} & \multicolumn{1}{c}{} & \multicolumn{1}{c}{} & \multicolumn{1}{c}{} &  \\ \hline
        \multicolumn{1}{c}{GU513} & \multicolumn{1}{c}{\textbf{1.26}} & \multicolumn{1}{c}{1.57} & \multicolumn{1}{c}{\textbf{0.99}} & \multicolumn{1}{c}{1.23} & \multicolumn{1}{c}{\textbf{16.02}} & 26.30 \\ \hline
    \end{tabular}
    \caption{Summary of foF2 error statistics for all stations using SSDMD and IRI.}
    \label{tab:fof2_error}
\end{table}

\begin{table}[h!]
    \centering
    \begin{tabular}{ccccccc}
        \hline
        \multicolumn{7}{c}{\textbf{hmF2 Forecast Errors}} \\ \hline
        \multicolumn{1}{c}{} & \multicolumn{2}{c}{RMSE} & \multicolumn{2}{c}{MAE} & \multicolumn{2}{c}{MAPE} \\ \hline
        \multicolumn{1}{c}{Station} & \multicolumn{1}{c}{SSDMD} & \multicolumn{1}{c}{IRI} & \multicolumn{1}{c}{SSDMD} & \multicolumn{1}{c}{IRI} & \multicolumn{1}{c}{SSDMD} & IRI \\ \hline
        \multicolumn{1}{c}{BC840} & \multicolumn{1}{c}{\textbf{23.03}} & \multicolumn{1}{c}{28.68} & \multicolumn{1}{c}{\textbf{16.41}} & \multicolumn{1}{c}{22.65} & \multicolumn{1}{c}{\textbf{6.72}} & 9.72 \\ \hline
        \multicolumn{1}{c}{} & \multicolumn{1}{c}{} & \multicolumn{1}{c}{} & \multicolumn{1}{c}{} & \multicolumn{1}{c}{} & \multicolumn{1}{c}{} &  \\ \hline
        \multicolumn{1}{c}{RO041} & \multicolumn{1}{c}{22.72} & \multicolumn{1}{c}{\textbf{22.15}} & \multicolumn{1}{c}{17.20} & \multicolumn{1}{c}{\textbf{16.80}} & \multicolumn{1}{c}{5.91} & \textbf{5.66} \\ \hline
        \multicolumn{1}{c}{} & \multicolumn{1}{c}{} & \multicolumn{1}{c}{} & \multicolumn{1}{c}{} & \multicolumn{1}{c}{} & \multicolumn{1}{c}{} &  \\ \hline
        \multicolumn{1}{c}{GA762} & \multicolumn{1}{c}{\textbf{38.30}} & \multicolumn{1}{c}{45.87} & \multicolumn{1}{c}{\textbf{30.32}} & \multicolumn{1}{c}{34.41} & \multicolumn{1}{c}{\textbf{13.08}} & 16.02 \\ \hline
        \multicolumn{1}{c}{} & \multicolumn{1}{c}{} & \multicolumn{1}{c}{} & \multicolumn{1}{c}{} & \multicolumn{1}{c}{} & \multicolumn{1}{c}{} &  \\ \hline
        \multicolumn{1}{c}{GU513} & \multicolumn{1}{c}{\textbf{45.43}} & \multicolumn{1}{c}{54.74} & \multicolumn{1}{c}{\textbf{34.00}} & \multicolumn{1}{c}{41.41} & \multicolumn{1}{c}{\textbf{9.84}} & 11.90 \\ \hline
    \end{tabular}
    \caption{Summary of hmF2 error statistics for all stations using SSDMD and IRI.}
    \label{tab:hmf2_error}
\end{table}

\section{Conclusions and Future Directions}\label{sec:conclusions}
We presented the standard DMD algorithm and formalized extensions that account for oscillations at multiple scales within measured data. Wavelet decompositions along each spatial dimension separated various scales within the time series that may otherwise appear as noise and will often preclude a standard DMD approach. For each of the scales, an affiliated reconstruction of the EDP time series was generated. Subsequent correlation analysis across the time scales then showed how we may recombine specific scales to preserve strong dynamic couplings between them in their one-step correlation. We called these correlated scales the connected components of the model. We performed an averaging step for each connected component by computing the mean over 24-hour time lags. This process denoises the data without erroneously removing oscillations from the original EDP signal that may initially appear as noise. Computing DMD on the connected components individually alleviates the problem of having large single-step gradients in the measurement data that would prevent DMD from fitting any stable modes. With each connected component, we produced a set of DMD eigenvalues and modes that summed coherently to form the SSDMD model. The final foF2 and hmF2 forecasts were then determined from the predicted EDPs.

SSDMD is one among many recent attempts to improve short-term forecasts of the foF2 and hmF2 parameters {\it cf.} \cite{perrone_2022, wang_2020, tsagouri_2018, mikhailov_2014, zhang_2013}. While other methods generally treat past foF2 or hmF2 measurements as inputs to the model, SSDMD instead uses the full EDP. It is the high-dimensionality of the EDP along with the use of DMD that gives our method a degree of dynamic expressivity that using the foF2 and hmF2 parameters alone would not.

The SSDMD algorithm is computationally efficient compared to physics-based models such as TIME-GCM or SAMI3, fitting a model and simulating a 5-minute resolution, 2-day forecast on the order of seconds using a single core on a consumer laptop. Therefore, SSDMD is lightweight enough to be updated in near-real-time as additional data are obtained, and it adapts to different measurement cadences without any changes to the model parameters. Additionally, SSDMD requires far less data to generate and update than empirical models like IRI or assimilation models like IRI-Real-Time-Assimilative-Mapping (IRTAM) \cite{galkin_2012} and the Global Assimilation of Ionospheric Measurements (GAIM) model \cite{shunk_2004}. With limited observations, as is the case with a single vertical ionosonde, SSDMD can produce reasonable forecasts of the average profile dynamics in the low, mid, and high latitudes. With high enough measurement cadence, the method should produce reliable short-term forecasts during periods of either solar maximum or solar minimum. A final added benefit of the SSDMD approach is the model has only four major hyperparameters, see Table \ref{tab:params_summary}, making it relatively simple to tune when necessary.

SSDMD fits a linear model to an expansion of EDP time series and thus may be seen as an autoregressive approach to forecasting foF2 and hmF2, and the simplicity of the approach makes it accessible to a wide range of operational and research applications. Still, the method is not without its limitations, as SSDMD does not account for any external driving forces such as solar activity, tidal forcing, or geomagnetic activity. As such, model forecast accuracy is highly dependent on there being strong correlations between the measurement and forecast periods at each time of day. Predicting anomalous events in the data is not possible without the inclusion of driving forces. Extending SSDMD further to incorporate external forcing is the topic of future development and, combined with longer measurement series, should allow for a significant increase in forecast accuracy. The DMD method can be modified to include control variables \cite{proctor_2016}, and in \cite{mehta_2018} a version of this method was implemented for a global model to great effect. Nevertheless, this model was fit using simulated data, whereas SSDMD aims to address the multiscale nature of measured EDPs. For this reason, the applicability of SSDMD to periods of prolonged or recurrent F-layer perturbations during quiet geomagnetic conditions may also be explored in future work. These disturbances can induce long-lived deviations in foF2 and hmF2 with magnitudes that far exceed climatology \cite{perrone_2020, zawdie_2020} which would not necessarily be captured by empirical models with drivers derived from geomagnetic and solar indices.

While the method was developed for one-dimensional observations of the ionosphere at a single sounder station, in future work, data from the global network of sounders may be used. However, a global model will require fitting additional spatial expansion functions to interpolate between the stations. Finally, data spanning longer time periods may also be used to extract seasonal and solar cycle dynamics. The method of SSDMD is ultimately not limited to ionospheric prediction, and it should be adaptable not only to other space weather domains, but many other systems that involve low-dimensional dynamics embedded in high-dimensional, multiscale observations.

\section*{Acknowledgements} \label{sec:acknowledgements}
This work was supported by the Naval Information Warfare Center Pacific (NIWC Pacific) and the Office of Naval Research (ONR).  

The authors would like to especially thank Dr. Douglas Drob from the Naval Research Laboratory Space Science Division for his insights and perspective on the use of this method. The authors would also like to extend thanks to Dr. Terry Bullet from the National Centers for Environmental Information, NOAA, and Dr. Ivan Galkin from the Lowell GIRO Data Center (LGDC) for the data they continue to collect and provide access to.

\section*{Data Availability Statement}
The code used in this study is openly available at \url{https://github.com/JayLago/SSDMD-Ionosphere}. The data used was obtained through the LGDC, \url{https://giro.uml.edu/didbase/}, using the SAO Explorer program, for which we are grateful to the developers and maintainers.

\appendix
\section{Koopman Mode Analysis}\label{sec:koopman}
Dynamic Mode Decomposition may be seen as a finite-dimensional approximation to the Koopman operator \cite{koopman}. The Koopman operator demonstrates how the equations for a generic nonlinear dynamical system may be rewritten as a linear infinite-dimensional operator acting on measurement functions of the system. This begins by considering a generic dynamical system,
\begin{equation}
    \frac{d}{dt}{\bf y}(t) = f({\bf y}(t)),  \quad {\bf y}(0) = {\bf y}_0 \in \mathcal{M} \subseteq \mathbb{R}^{N_{s}},
\end{equation}
where $\mathcal{M}$ is some connected, compact subset of $\mathbb{R}^{N_{s}}$ and define an {\it observable}, $g({\bf y}(t))$,  such that $g: \mathcal{M} \mapsto \mathbb{C}$.  Denoting the affiliated flow, ${\bf y}(t) = S(t; {\bf y})$, we may rewrite the system using the {\it Koopman operator}, $\mathcal{K}^{t}$, 
\begin{equation} \label{eqn:koop1}
    \mathcal{K}^{t}g({\bf y}) =  g\left(S(t;{\bf y})\right).
\end{equation}
We see $\mathcal{K}^{t}$ is linear since
\begin{eqnarray}
    \mathcal{K}^{t}(\alpha g_1({\bf y}) + \beta g_2({\bf y})) & = & \alpha g_1(S(t; {\bf y})) + \beta g_2(S(t; {\bf y}))  \nonumber \\
    & = & \alpha \mathcal{K}^{t} g_1({\bf y}) + \beta \mathcal{K}^{t} g_2({\bf y}).
\end{eqnarray}
Following \cite{lago_2022},  we see that with some basic assumptions, i.e. if we choose observables such that they are square-integrable and suppose $\mathcal{M}$ is invariant with respect to the flow, we have simplified a problem of determining some unknown nonlinear function $f({\bf y}(t))$ to one of finding an eigendecomposition of the linear operator, $\mathcal{K}^{t}$.  Moreover, by finding the Koopman eigenfunctions
\begin{equation}
    \{\phi_j\}_{j=1}^{\infty}
\end{equation}
and affiliated eigenvalues
\begin{equation}
    \{\lambda_j\}_{j=1}^{\infty},
\end{equation}
where
\begin{equation}
    \mathcal{K}^{t}\phi_{j} = e^{t\lambda_{j}}\phi_{j}, \quad j\in \{1,2,\dots \},
\end{equation}
then we have a modal decomposition for any other observable, $g$, so that
\begin{equation}
    g({\bf y}) = \sum_{j=1}^{\infty}c_{j}\phi_{j}({\bf y}),
\end{equation}
and we can track the evolution of $g({\bf y})$ along the flow with the formula,
\begin{equation} \label{eqn:koop2}
    \mathcal{K}^{t}g({\bf y}) = \sum_{j=1}^{\infty}c_{j} e^{t \lambda_{j}} \phi_{j}({\bf y}).
\end{equation}
See \cite{budisic} and \cite{mezic4} for more in-depth treatments of the Koopman operator and its properties, \cite{mezic1, kutz} for deeper connections between DMD and Koopman, and \cite{schmid, tu1, williams} for additional details on the DMD algorithm and its variations. We point out that the Koopman operator is most naturally formulated with respect to Lagrangian data while in this work we focus on analyzing Eularian data, that is to say, we assume the ${\bf y}_j$ observations in our data stream are measurements of the EDP at fixed positions in altitude. Were one to develop effective Euler-to-Lagrangian maps for the data sets studied herein, this would open up a wider range of tools related to the DMD method. This is a subject for future research.

\section{Pseudocode Algorithm}\label{sec:algorithm}
The complete SSDMD method is summarized in Algorithm \ref{alg:ssdmd}. We assume familiarity with standard numerical methods for computing the reduced Singular Value Decomposition (SVD), eigenvalue decomposition, solving an initial value problem, and computing 1-dimensional wavelet decompositions. When computing the mean profiles over 24-cycles, use Equation \ref{eqn:meancycles}. The algorithm returns the reconstructed time series of the input data along with the DMD eigenvalues, modes, and eigenfunctions.
\begin{algorithm}
\DontPrintSemicolon
\KwData{${\bf Y} \in \mathbb{R}^{N_{S} \times N_{T}}$ such that each column, ${\bf y}_i \in \mathbb{R}^{N_{S}}$, is an observation of the system $\delta t$ time from ${\bf y}_{i-1}$.}
\KwResult{$\hat{\bf Y}, {\bf W}, \bm{\Lambda}, {\bf \Phi}$}
Initialize: set DMD threshold $c_{\text{dmd}}>0$, and correlation threshold $c_{\text{corr}}>0$.\;
\Begin{
	$\tilde{\bf Y} \longleftarrow discreteWaveletDecomposition({\bf Y})$\;
	$\tilde{\bf Y}^{\text{C}}, N_{\text{C}} \longleftarrow correlatedConnectedComponents(\tilde{\bf Y}, c_{tr})$\;
	\For{n=1 \dots $N_{\text{C}}$}{
		$\bar{\bf Y}_{n}^\text{C} \longleftarrow meanDailyCycles(\tilde{{\bf Y}}_{n}^\text{C})$\;
		$\bar{{\bf Y}}_{n,-}^\text{C} \longleftarrow \left[\bar{{\bf y}}_{n,1}^\text{C}~\bar{{\bf y}}_{n,2}^\text{C}~\cdots\bar{{\bf y}}_{n,m-1}^\text{C} \right]$\;
		$\bar{{\bf Y}}_{n,+}^\text{C} \longleftarrow \left[\bar{{\bf y}}_{n,2}^\text{C}~\bar{{\bf y}}_{n,3}^\text{C}~\cdots\bar{{\bf y}}_{n,m}^\text{C} \right]$\;
		${\bf U}, \bm{\Sigma}, {\bf V}^{\dagger} \longleftarrow reducedSVD(\bar{{\bf Y}}_{n, -}^{\text{C}}, C_{dmd})$\;
		${\bf K}  \longleftarrow \bar{{\bf Y}}_{n,+}^{\text{C}} {\bf V} \bm{\Sigma}^{-1} {\bf U}^{\dagger}$\;
		${\bf W}_{n},  \bm{\Lambda}_{n} \longleftarrow eigenvalueDecomposition({\bf K})$\;
		${\bf \Phi}_{n} \longleftarrow solveIVP({\bf W}_{n},\bar{{\bf Y}}_{n,-}^{\text{C}})$\;
		${\hat{\bf Y}}_{n} \longleftarrow {\bf W}_{n}{\bf \Lambda}_{n}{\bf \Phi}_{n}$\;
	}
	$\hat{\bf Y} \longleftarrow \sum_{n=1}^{N_{\text{C}}} \hat{\bf Y}_{n} $\;
	${\bf W} \longleftarrow \left[{\bf W}_{1}~{\bf W}_{2}~\cdots~{\bf W}_{n}\right]$\;
	${\bf \Lambda} \longleftarrow \left[{\bf \Lambda}_{1}~{\bf \Lambda}_{2}~\cdots~{\bf \Lambda}_{n}\right]$\;
	${\bf \Phi} \longleftarrow \left[{\bf \Phi}_{1}~{\bf \Phi}_{2}~\cdots~{\bf \Phi}_{n}\right]$\;
}
\caption{SSDMD\label{alg:ssdmd}} 
\end{algorithm}

\clearpage
\bibliography{ms}
\bibliographystyle{unsrt}
\end{document}